\documentclass[sigconf, balance=False]{acmart}
\usepackage{popets}

\setcopyright{popets}
\copyrightyear{YYYY}

\acmYear{YYYY}
\acmVolume{YYYY}
\acmNumber{X}
\acmDOI{XXXXXXX.XXXXXXX}
\acmISBN{}
\acmConference{Proceedings on Privacy Enhancing Technologies}
\usepackage{algorithmic}
\usepackage{algorithm}
\usepackage{graphicx}
\begin{document}

\title{DeTorrent: An Adversarial Padding-only Traffic Analysis Defense}


\author{James K Holland}
\affiliation{%
  \institution{University of Minnesota}
  \city{Minneapolis}
  \state{}
  \country{USA}}
\email{holla556@umn.edu}

\author{Jason Carpenter}
\affiliation{%
  \institution{University of Minnesota}
  \city{Minneapolis}
  \state{}
  \country{USA}}
\email{carpe415@umn.edu}

\author{Se Eun Oh}
\affiliation{%
  \institution{Ewha Womans University}
  \city{Seoul}
  \state{}
  \country{South Korea}}
\email{seoh@ewha.ac.kr}

\author{Nicholas Hopper}
\affiliation{%
  \institution{University of Minnesota}
  \city{Minneapolis}
  \country{USA}}
\email{hoppernj@umn.edu}

\renewcommand{\shortauthors}{Holland et al.}

\begin{abstract}

While anonymity networks like Tor aim to protect the privacy of their users, they are vulnerable to traffic analysis attacks such as Website Fingerprinting (WF) and Flow Correlation (FC). Recent implementations of WF and FC attacks, such as Tik-Tok and DeepCoFFEA, have shown that the attacks can be effectively carried out, threatening user privacy. Consequently, there is a need for effective traffic analysis defense.

There are a variety of existing defenses, but most are either ineffective, incur high latency and bandwidth overhead, or require additional infrastructure. As a result, we aim to design a traffic analysis defense that is efficient and highly resistant to both WF and FC attacks. We propose DeTorrent, which uses competing neural networks to generate and evaluate traffic analysis defenses that insert `dummy' traffic into real traffic flows. DeTorrent operates with moderate overhead and without delaying traffic. In a closed-world WF setting, it reduces an attacker's accuracy by 61.5\%, a reduction 10.5\% better than the next-best padding-only defense. Against the state-of-the-art FC attacker, DeTorrent reduces the true positive rate for a $10^{-5}$ false positive rate to about .12, which is less than \textit{half} that of the next-best defense. We also demonstrate DeTorrent's practicality by deploying it alongside the Tor network and find that it maintains its performance when applied to live traffic. 

\end{abstract}


\maketitle

\footnotetext[1]{James K Holland and Se Eun Oh are the corresponding authors.}

\section{Introduction}

As people's personal and professional lives increasingly depend on the Internet, the threat of network surveillance has become particularly salient. To combat this threat, many users utilize privacy-enhancing tools such as Tor to protect themselves. Accordingly, Tor has surged in popularity, recording millions of users each day \cite{metrics, measurement}. Tor operates by encrypting traffic and routing it through a series of volunteer-run relays, ensuring that no single entity can observe both entry and exit traffic \cite{Dingledine2004}. The encryption strategy, known as onion routing, is characterized by messages encapsulated by multiple layers of encryption. Each successive Tor relay then strips a layer of encryption from the message and forwards it to the revealed destination \cite{onions}. 

While data sent through Tor may be secure, the frequency and timing of the associated network traffic may be used by traffic analysis techniques. These techniques include website fingerprinting \cite{Herrmann2009, Panchenko2011, Wang2013, Cai2014, Wang2014a, Hayes2015, Abe2016, Wang2016, Panchenko2017, Rimmer2018, Sirinam2018, Rahman2019, Bhat2019, triplet, Oh2019} and end-to-end flow correlation \cite{Raymond2001, Levine2004, ye2005, schmatikov2006, raptor, deepcorr, dcf}, which have been thoroughly discussed in past literature. Researchers have demonstrated that both can be carried out effectively. 

In a website fingerprinting (WF) attack, an adversary between the user and the first Tor relay records the timing and volume information of the Tor traffic and determines which website the user is visiting. Because this attack requires only a passive, local adversary, it is particularly threatening to user privacy. Alternatively, in the flow correlation (FC) attack setting, the adversary records traffic metadata as it both enters and exits the Tor network. Then, the adversary attempts to correlate associated entry and exit flows to determine both the origin and destination of the traffic. This attack is explicitly mentioned in early literature; however, security against end-to-end attacks is difficult for low-latency anonymity networks and an explicit non-goal of the original Tor design \cite{Dingledine2004}. 

A variety of traffic analysis defenses have been proposed to prevent website fingerprinting and flow correlation \cite{Wright2009, Luo2011, Cai2012, Cai2014, Juarez2015, Wang2017, Gong2020, banp, holland2022, dolos, machines, surakav}. These defenses typically operate by either delaying traffic or by padding with `dummy' traffic, obfuscating the relationship between the defended traffic and the associated flow. However, many defenses require either additional infrastructure or a prohibitive amount of latency or bandwidth overhead. While padding-only defenses may still incur some delay when widely deployed on Tor \cite{paddingonly}, we avoid explicitly delaying packets, as excessively increasing Tor latency is discouraged by Tor developers \cite{padding}. Accordingly, we aim to design an effective website fingerprinting and flow correlation defense that uses \textit{only} dummy traffic.  

The most recent and effective website fingerprinting and flow correlation attacks use deep neural networks (DNNs) to classify and link Tor traffic. DNNs are vulnerable to adversarial examples \cite{dnn}, which are intentionally crafted inputs that `trick' a DNN into misclassifying that input. As a result, it may appear that adversarial techniques could be used to design a traffic analysis defense. However, adapting adversarial techniques for traffic analysis is not as straightforward as it may initially appear. First, the defense must operate while the traffic is being sent and without knowledge of future traffic, while the attacker can collect the relevant data for later analysis. Second, the attacker will likely have access to the defense and can re-train the attack on a representative `defended' dataset. In other words, the attacker model is not a static model targetable by adversarial perturbations; instead, it can adapt and retrain based on Tor's choice of traffic analysis defense. Thus, the defense must act in an unpredictable manner that resists adversarial retraining. Previous work on using adversarial methods for traffic analysis defense has achieved a degree of success \cite{surakav, Rahman2021, banp}. However, these defenses either require knowledge of future traffic, struggle to prevent adversarial retraining, or delay traffic. Finally, the defense needs to be attack model-agnostic. For example, a defense model tailored to resist DNN-based attacks may not remain robust against alternate approaches, such as support vector machines. 

To overcome these challenges, we present a novel approach, DeTorrent, to generate traffic padding strategies that defend Tor traffic. Inspired by generative adversarial networks (GANs) \cite{gan}, the core of the defense consists of a pair of competing neural networks that aim to either disguise or identify Tor traffic. More specifically, the `generator' takes random noise as input and then schedules when and how many dummy packets should be sent. Then, the `discriminator' (or attacker) attempts to demonstrate that it can accurately identify the defended Tor traffic. Because the generator's loss function is minimized when preventing the attacker from being able to identify the traffic, it is incentivized to generate effective padding strategies. Most importantly, those strategies are designed for resistance to adversarial retraining, as the generator and discriminator are trained in tandem. 

Furthermore, to simulate the generator's real-time decision-making, we implement the generator as a long short-term memory (LSTM) neural network \cite{lstm}. An LSTM is a type of recurrent neural network (RNN) that can be used to make repeated decisions while maintaining a state to store historical information. In the defense, the LSTM outputs how many dummy packets should be sent at each time step. Because the LSTM has feedback connections, it can generate this padding pattern while considering the previously monitored traffic. Once the generator is trained, it can be used to make effective real-time dummy padding insertion strategies.

To test the DeTorrent defense, we evaluate its ability to reduce the accuracy of state-of-the-art website fingerprinting and flow correlation attacks. We find that it outperforms comparable defenses in both types of attacks while using similar bandwidth overhead. To illustrate, DeTorrent reduced closed-world Tik-Tok attack accuracy from 93.4\% to 31.9\%, which is 10.5\% better than the next best defense. Additionally, DeTorrent reduces DeepCoFFEA's true positive rate to .12 for a $10^{-5}$ false positive rate, which is less than half of Decaf's TPR at .29. It does this while adding a bandwidth overhead of 98.9\% in the WF setting and 97.3\% in the FC setting. 

To make our approach deployable, we provide a full implementation of the DeTorrent defense as a Tor pluggable transport and evaluate its performance against Tor live traffic. We further show the transferability of DeTorrent's padding strategies by training on one dataset partition and testing on another, leading to a minimal drop in performance of only .7\%. Overall, DeTorrent demonstrates that a reasonable degree of traffic analysis defense can be achieved while only adding dummy traffic. 

\section{Background and Related Work}

\subsection{Website Fingerprinting}

\begin{figure}
  \centering
  \includegraphics[scale=.7]{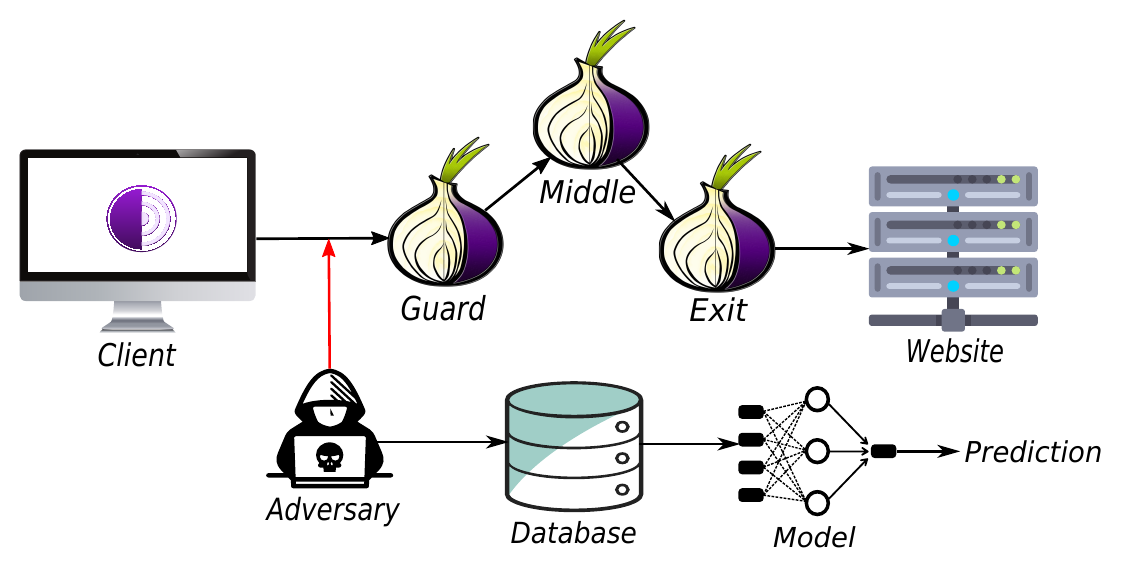}
  \caption{Attacker position and model creation for a WF attack on Tor. The adversary creates a database of download traces for multiple sites and uses them to train a model, which is applied to observed traces between a client and guard relay to predict the downloaded site.}
  \label{fig:tor_wf}
\end{figure}

\begin{figure}
  \centering
  \includegraphics[scale=.6]{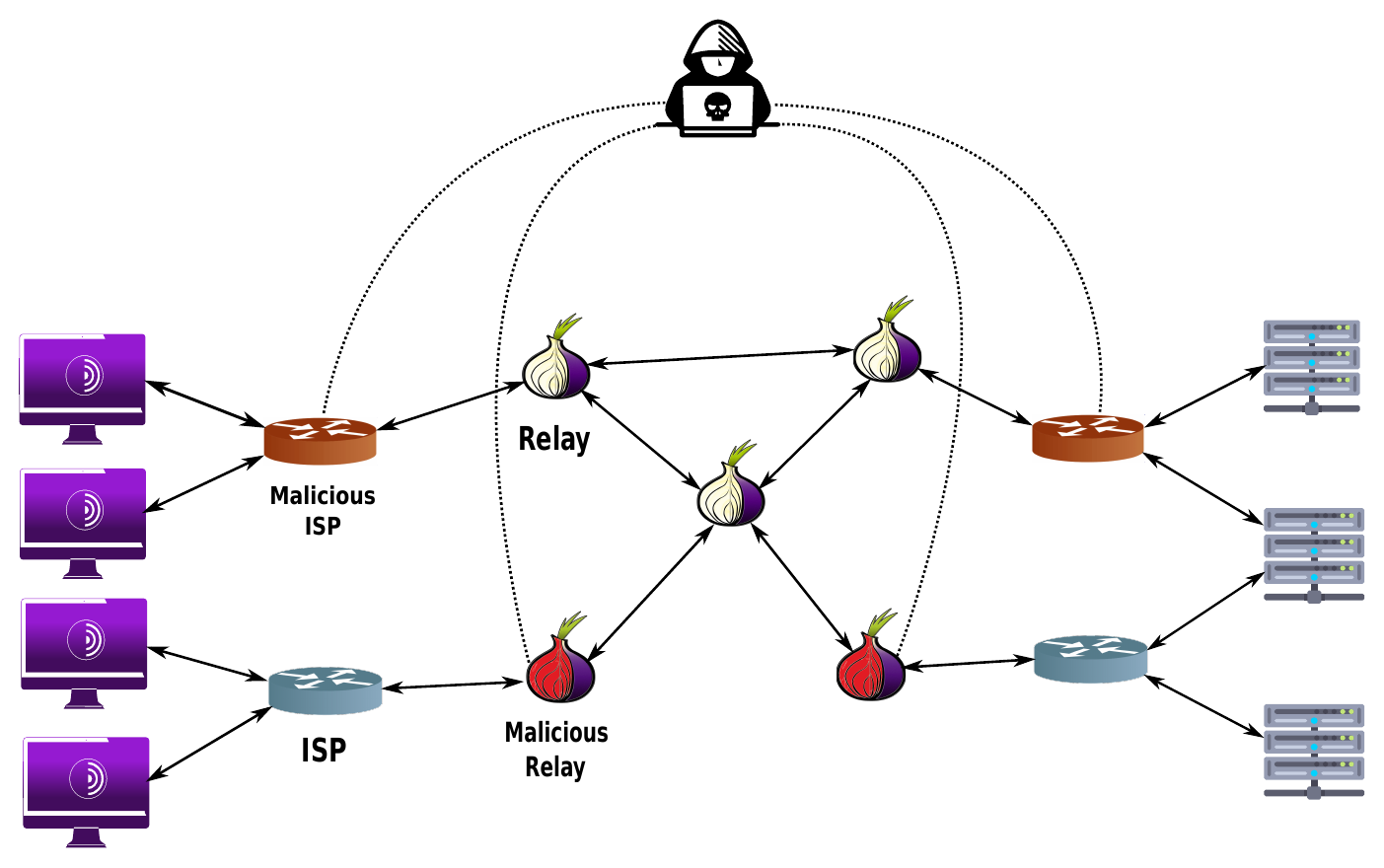}
  \caption{Attacker positioning for a FC attack on Tor. The attacker must collect traffic at both the entry and exit sides of the network, though there are multiple possible vantage points, including the ISPs and malicious relays.}
  \label{fig:tor_fc}
\end{figure}

\subsubsection{Threat Model}
In the WF attack model, shown in Figure \ref{fig:tor_wf}, we assume that the adversary is a \textit{passive} eavesdropper \textit{local} to the client. Accordingly, this adversary can observe traffic between the Tor client and the entry node but does not modify that traffic. We also assume that this adversary knows the IP address of the user and is trying to determine which website the user is visiting. 

\subsubsection{Attacks}

While data sent through Tor is encrypted, traffic metadata such as packet timing and direction may indicate the destination of a user's traffic. Hintz et al. were the first to demonstrate the risk of WF attacks using file transfer sizes to distinguish web pages \cite{Hintz2002}. Tor was initially thought to be WF-resistant due to its use of constant-sized cells and inherent network delays. However, researchers later began using packet volume and timing metadata to conduct the attack against Tor traffic \cite{Cai2012, Panchenko2011}. Later improvements came in the form of dataset collection and feature engineering \cite{Wang2013}, the use of more effective machine learning techniques such as k-nearest neighbors and random forests \cite{Wang2014a, Hayes2015}, and the use of the cumulative representation of the traces \cite{Panchenko2017}. 

More recent techniques leveraged deep neural networks to substantially increase WF attack performance \cite{Sirinam2018, Rahman2019, Oh2019, triplet}. Most importantly, Deep Fingerprinting by Sirinam et al. \cite{Sirinam2018} managed to defeat the WTF-PAD defense while pushing closed-world accuracy above 98\%. Then, Rahman et al. \cite{Rahman2019} added timing information to the Deep Fingerprinting architecture to develop the Tik-Tok attack, further increasing performance with especially strong results against the Walkie-Talkie defense \cite{Rahman2019}. Due to the success of the convolutional neural network architecture used by Deep Fingerprinting, we use it as the basis of DeTorrent's embedder and discriminator, with some modifications described in Appendix \ref{ssec:architecture}. While some recent attacks focused on outside sources of information to reduce false positives \cite{oracle} or training WF attacks in more challenging and data-limited settings \cite{triplet, gandalf}, we use the Deep Fingerprinting and Tik-Tok attacks to benchmark our defenses as they represent versatile state-of-the-art attacks that have proven to be effective against WF defenses. 

\subsubsection{Defenses}

WF defenses alter web traffic to reduce the attacker's ability to identify the website a user is visiting. Some early defense designs, such as BuFLO and Tamaraw, operate by sending packets at a set rate and inserting dummy packets when no real data is available, making traffic associated with different websites look identical \cite{Dyer2012, Cai2014a, Cai2014}. However, these defenses incur prohibitively high latency and bandwidth overheads. Another defense, RegulaTor \cite{holland2022}, also aims to regularize traffic to prevent accurate classification. However, it operates with a more reasonable overhead by sending real traffic `surges' in standardized bursts, decreasing the amount of information leaked. Because these defenses delay traffic, they are not used for comparison in this paper. 

Other defenses operate by transforming traffic such that it appears to be associated with a different website or groups of websites, thus tricking the WF attacker. These techniques include Traffic Morphing \cite{Wright2009}, Decoy pages \cite{Panchenko2011}, Supersequence \cite{Wang2014a}, and Glove \cite{Nithyanand2014}. Additionally, the Walkie-Talkie defense \cite{Wang2017} modifies the browser to operate in half-duplex mode while modifying burst sequences. These defenses have either been broken by WF attacks or require additional infrastructure or information about traffic associated with other web pages, so we do not evaluate them in this paper. 

Additional latency overhead is harmful to the user experience, so some defenses have taken the approach of avoiding traffic delays. WTF-PAD \cite{Juarez2015} uses adaptive padding and distributions of typical inter-packet delays to fill the large gaps in traffic that may leak information about the trace destination. FRONT \cite{Gong2020}, on the other hand, focuses dummy padding on the beginning of the trace while randomizing dummy packet volume and timing information. Additionally, the Spring and Interspace \cite{machines} defenses operate using the Tor circuit padding framework \cite{padding}, which defines state machines controlling when dummy traffic should be inserted. The state machines are designed using genetic algorithms with a fitness function based on the defense's ability to defend against the Deep Fingerprinting attack. Because these defenses have reasonable overhead with no added latency, we directly compare them to DeTorrent. 

Some defenses use adversarial techniques to defend traffic. Mockingbird \cite{Rahman2021} generates adversarial traces that aim to reduce the attacker's adversarial retraining accuracy by randomly perturbing in the space of viable traces. While this approach appears promising, it cannot be implemented live, as it requires knowledge of future traffic. Shan et al. \cite{dolos} designed a defense, Dolos, that precomputed input-agnostic `adversarial patches' that could be used to disrupt deep learning-based attacks in real-time. Nasr et al. \cite{banp} proposed a technique using blind adversarial perturbations (BAP) to defeat DNN-based traffic analysis. Essentially, the approach trained DNN adversarial traffic generators that aimed to reduce the performance of WF and FC attacks. While BAP is quite effective against a WF or FC adversary using a static DNN model, we assume in this paper that the attacker will be able to retrain their models once traffic is defended. Accordingly, our BAP implementation in this paper will likely appear less effective than expected. Still, BAP's use of a DNN traffic perturbation generator was influential to DeTorrent's development. Surakav \cite{surakav} uses a Wasserstein GAN \cite{wasserstein} to generate traffic burst volumes for a trace, where a burst is defined as a series of packets going in the same direction. Then, Surakav shapes the real traffic bursts to match that of the generated traffic. While Surakav appears effective, it delays packets while shaping the bursts, so it is not included in our analysis. 

The TrafficSliver \cite{trafficsliver} and Multihoming \cite{multihoming} defenses split traffic among different guard nodes such that the individual sub-traces are difficult for a WF attacker to classify. Specifically, Multihoming users connect through different access points (such as Wi-Fi and a cellular network) and merge the traffic at a multipath-compatible Tor bridge. Alternatively, TrafficSliver either splits traffic over multiple entry relays and merges traffic at the middle relay or creates multiple circuits and requests different HTTP objects over the circuits. While these defenses appear effective, they do not guard against all attacker types considered in this paper. For example, TrafficSliver does not defend against local attackers who can see outgoing traffic before it is split. Furthermore, Multihoming requires additional infrastructure to function. As a result, traffic-splitting defenses are not used for comparison in this paper. 

\subsection{Flow Correlation}


Early discussion of security concerns in low-latency anonymity networks acknowledged the possibility of timing analysis to correlate flows, determine if two mixes are on the same path, or simply confirm that two users are online at the same time  \cite{Raymond2001, Wang2002, Levine2004, Dingledine2004, murdoch2005, overlier2006}. While initial attack formulations included techniques such as `packet counting' \cite{Raymond2001}, later approaches turned to statistical correlation metrics such as the cross-correlation between binned packet volumes \cite{schmatikov2006, Levine2004}, distance functions using mutual information and frequency analysis \cite{ye2005}, or Spearman's rank correlation coefficient \cite{raptor}. Defensive techniques proposed in response include defensive dropping \cite{Levine2004}, adaptive padding \cite{schmatikov2006}, and constant-rate sending \cite{Levine2004}. However, constant-rate sending is costly in terms of bandwidth and latency overhead, while defensive dropping has been proven ineffective \cite{schmatikov2006}. Still, we test the performance of an adaptive padding approach in the flow correlation setting later in this paper. 

The original Tor design admits vulnerability to end-to-end timing correlation, though the attacker must watch traffic at both the initiator and responder to carry this out \cite{Dingledine2004}. Later work introduced the concept of an autonomous system (AS) level adversary \cite{feamster2004}. While these adversaries have a more limited view of the network, it is still quite possible for multiple nodes to be part of the same administrative domain. This risk is then exacerbated by the existence of Internet exchange points (IXPs), which allow the exchange of traffic between autonomous systems, as well as the possibility of IXP coalitions \cite{johnson2013}. AS-level adversaries may also be able to increase their chances of intercepting traffic through active attacks, such as BGP hijacks and BGP interception \cite{raptor, Tan2016}. As a result, reducing the chance that an adversary is positioned to capture both entry and exit flows is a concern for Tor researchers and developers. In response, several research efforts address these problems by improving Tor's guard and relay selection \cite{edman2009, Nithyanand2014, akhoondi2012, juen2014, sun2017, Armon2016}. 

A related threat model is the possibility of an adversary running multiple Tor relays with the expectation that a targeted client will eventually use a controlled relay as both the entry and exit. This adversary may improve their odds by preventing the creation of circuits that they're a part of, but cannot yet compromise \cite{Borisov07}. Currently, the Tor Project's response is to have users re-use entry guards for a set time period to reduce the cumulative risk of eventually choosing a compromised circuit \cite{rotation}. See Figure \ref{fig:tor_fc} for an illustration of a flow correlation attack that can make use of either relays or internet infrastructure. 

Even if an adversary can capture entry and exit flows, correlating them is nontrivial due to Tor's use of fixed-sized cells and varying circuit conditions that may alter flow characteristics. Furthermore, the base rate fallacy is a concern while carrying out a flow correlation attack due to the number of potential flow combinations: to deanonymize a set of $N$ entry and exit flows, $N^2$ possible combinations of flows must be evaluated. This is both computationally and practically difficult, as the number of false positive correlations will likely outnumber the number of true correlations unless the false positive rate is relatively low. 

However, these challenges have been overcome by recent work by Nasr et al. and Oh et al. with their DeepCorr and DeepCoFFEA attacks \cite{deepcorr, dcf}. DeepCorr first used a deep neural network to learn how Tor circuits transform traffic, correlating flows with a lower false positive rate while using less data than previous attacks. DeepCoFFEA provides further improvements by training feature embedding networks to represent the entry and exit flows in a low-dimensional space such that correlated flows are close to one another. DeepCoFFEA also further reduces the false positive rate by dividing flows into windows and voting across the windows where the window embeddings are similar. These attacks essentially demonstrate that flow correlation is quite feasible for attackers to effectively carry out given the ability to intercept entry and exit flows. Given that DeepCoFFEA is a state-of-the-art flow correlation attack, we use it in this paper to test the performance of various potential flow correlation defenses. We also implemented the flow correlation defense Decaf, which was presented alongside the DeepCoFFEA attack. It operates by recording windows of traffic from other traffic flows, and then using the packet timing from a randomly selected window to defend a window of real traffic. This way, the window of real traffic is more difficult to embed accurately. 

\subsection{LSTMs}

Recurrent neural networks are a class of artificial neural networks designed for processing sequential data. Unlike traditional feedforward networks, RNNs have internal loops that allow information to persist, enabling them to use previous steps' outputs as inputs for the current step. While RNNs can theoretically consider longer-term input dependencies in sequential data, they suffer from the `vanishing gradient' problem, where back-propagated gradients tend to either zero or infinity. To address this, LSTMs add a set of `gates,' called the input, output, and forget gates, that regulate information across cell states. Consequently, the LSTM is better able to consider long-term dependencies, giving it a `long short-term memory.' LSTMs also support the use of irregularly-spaced time series of variable length, as they can make repeated decisions throughout the time series. As a result, an LSTM can make a series of decisions regarding inserting dummy packets into a traffic flow \textit{while the traffic is being sent}. 

\section{DeTorrent Design}

\subsection{Motivation}

A major difficulty faced by website fingerprinting defenses is that the attacker likely has access to the defense and defended traffic. Consequently, they can retrain their models or improve their approach to undermine the defense more effectively. To defeat this adversarial training, the most effective defenses have utilized the padding strategies based on a high degree of randomness as demonstrated by FRONT \cite{Gong2020}, Surakav \cite{surakav}, and Mockingbird \cite{Rahman2021}. 

As a result, it may seem intuitive to use an approach similar to that of a Generative Adversarial Network (GAN) to generate the defended traces and make adversarial retraining less effective. In this setup, the generator is fed random noise as input and trained to maximize the discriminator's loss. This way, the generator learns to insert dummy traffic in a manner that minimizes the discriminator's ability to identify the underlying traffic. However, this approach yields three challenges: representing the trace in a manner that can be used by neural networks, training the generator to make decisions about scheduling dummy traffic, and providing a notion of how well the discriminator can identify the trace. Our approaches for the FC and WF settings vary somewhat, so we now refer to the website fingerprinting version of DeTorrent as WF-DeTorrent and the flow correlation version as FC-DeTorrent. 

\subsection{Trace Representation}
\label{trace_representation}
Representing the trace as a tensor is difficult given that the representations must support being `added' to one another while serving as the output of one neural network and the input of another. This problem is further complicated by the fact that the LSTM generator must output the dummy padding pattern through a series of time steps. Also, because the generator's loss function is written in terms of minimizing the discriminator's performance, it must alter the traffic representation in a differentiable manner in order to use backpropagation to update its weights. In other words, we must be able to backpropagate through the discriminator's output and back to the generator's weights. This prevents us from representing the traffic as a series of packet directions or timestamps, as shown in existing WF attacks such as Deep Fingerprinting \cite{Sirinam2018}.

We find the best approach to account for these difficulties is to bin the packets of the trace such that the value of each bin is the volume of download packets sent during a designated time period. This way, the traces can be `defended' by simply adding the tensors together. This representation only considers download packets because the upload traffic is generally both smaller in volume and distributed similarly compared to the download traffic, as discussed in past work \cite{holland2022}. Thus, the upload padding is handled based on observed download traffic, as discussed in Section \ref{ssec:packet_timing}. Because a majority of traffic is sent soon after a web page is loaded, the bins are spaced evenly on a logarithmic scale using the NumPy function `geomspace' \cite{geomspace}, allowing for more fine-grained distinction between early traffic bursts. However, the bins are only fit to the data in the sense that they span 50 seconds, which is enough to contain the traffic in nearly all web page loads. The timing for each bin is also consistent for all traces. As a result, bin spacing will most likely not need to be re-tuned even if web browsing traffic patterns change. 

While a binned representation loses some fine-grained information about the traces, this appears to be minimal compared to the amount of information leaked by the overall distribution and volume of traffic. We test this by running a closed-world WF classification attack on the DF dataset using only the binned representations and find that accuracy remains high at over 90\%. This is further suggested by the success of WF attack CUMUL \cite{Panchenko2017}, which uses the cumulative distribution traces, and the relative success of WF defenses, such as FRONT, RegulaTor, and Surakav, that primarily alter the volume and distribution of traffic \cite{Gong2020, holland2022, surakav}. Furthermore, the success of the recent Robust Fingerprinting attack, which uses time slots, or bins, to create a `robust' traffic representation, is further evidence of the usefulness of this representation \cite{robust}. 

Another potential limitation of the binning is that the DeTorrent generator is unable to learn the precise timing of dummy traffic and instead must rely on heuristic strategies, as described in Section \ref{ssec:packet_timing}. However, a more fine-grained version of DeTorrent would likely also struggle to respond to observed real traffic quickly enough to be effective, as this would require much more frequent LSTM evaluations. Having smaller bins that the beginning of traces also helps to mitigate this somewhat. 

To preprocess the data, we first shift the traffic so that the binning starts with the 10th packet, as the time for the website to begin loading is variable. However, we can prevent this variability from impacting the trace representations by `subtracting out' the time it takes for the circuit to be constructed and for the web page to begin loading. When DeTorrent is applied to real traffic, it sends additional dummy traffic with inter-packet delays drawn from an exponential distribution with a mean delay of one-tenth of a second. This obfuscates the circuit setup and initial communication with the website while allowing the real traffic to be sent without delay. 

\subsection{Trace Embedding}

\label{ssec:trace_embedding}
A naive implementation of WF-DeTorrent would create loss functions based on the discriminator's ability to classify traces (or correlate flows) and the generator's ability to increase the discriminator's loss. However, this style of training does not lead to an effective defense in the WF setting. This is because the generator is incentivized to defend the traffic in a manner that does not resist retraining, as it is easy to move the trace representation just across the WF discriminator's decision boundary and cause it to misclassify. However, the discriminator can retrain to account for the generator's minor perturbations. To provide a better metric for discriminator and generator performance in the WF setting, we chose an approach used by DeepCoFFEA \cite{dcf} and trained an embedder to map traces to a low-dimensional space where trace similarity is measured using Euclidean distance. 

To train the embedder to provide a useful notion of distance between traces, we use triplet loss, which compares an `anchor' (A) input of a given class to both a `positive' (P) input of the same class and a `negative' (N) input from another class. Triplet loss is minimized when the distance between the anchor and positive embeddings is at least $\alpha$ smaller than the distance between the anchor and negative embeddings, where $\alpha$ is a set slack value. \\

$\mathcal{L}(A,P,N) = max(||f(A) - f(P)||^2 - ||f(A) - f(N)||^2 + \alpha, 0)$
\\\\
Once the embedder is trained, the discriminator guesses the embedding of the original traffic while the generator defends the traffic to prevent this. By making accurate guesses, the discriminator indicates that it has enough information to identify the traffic. As discussed later, we find that an embedding dimension of 256 works best when considering defense performance and computational overhead. In the flow correlation setting, the discriminator identifies whether the flow pairs are matched or not. Because this is a binary classification, the problem of short-sighted optimization does not occur, so embedding is not needed. 

\subsection{Real-Time Decisions}

While a WF or FC defense could pre-generate padding strategies, it would likely underperform compared to a similar defense that takes into account traffic already sent or received during page load. For example, a real-time defense might decide to extend packet bursts or add fake traffic to traces with low packet volume. Similarly, the generator cannot take the entire trace as input, as the real-world implementation lacks knowledge of traffic that has not yet been sent. To train a generator to make these types of decisions, we implement it as an LSTM where each bin in the trace representation is matched to a time step of the LSTM. As a result, the number of decisions the LSTM makes is tied to the length of the trace representation, though this representation is of variable length. This way, the recurrent nature of the LSTM can be used to make repeated decisions while taking into account past traffic and padding decisions. 

\begin{figure}
  \centering
  \includegraphics[scale=.62]{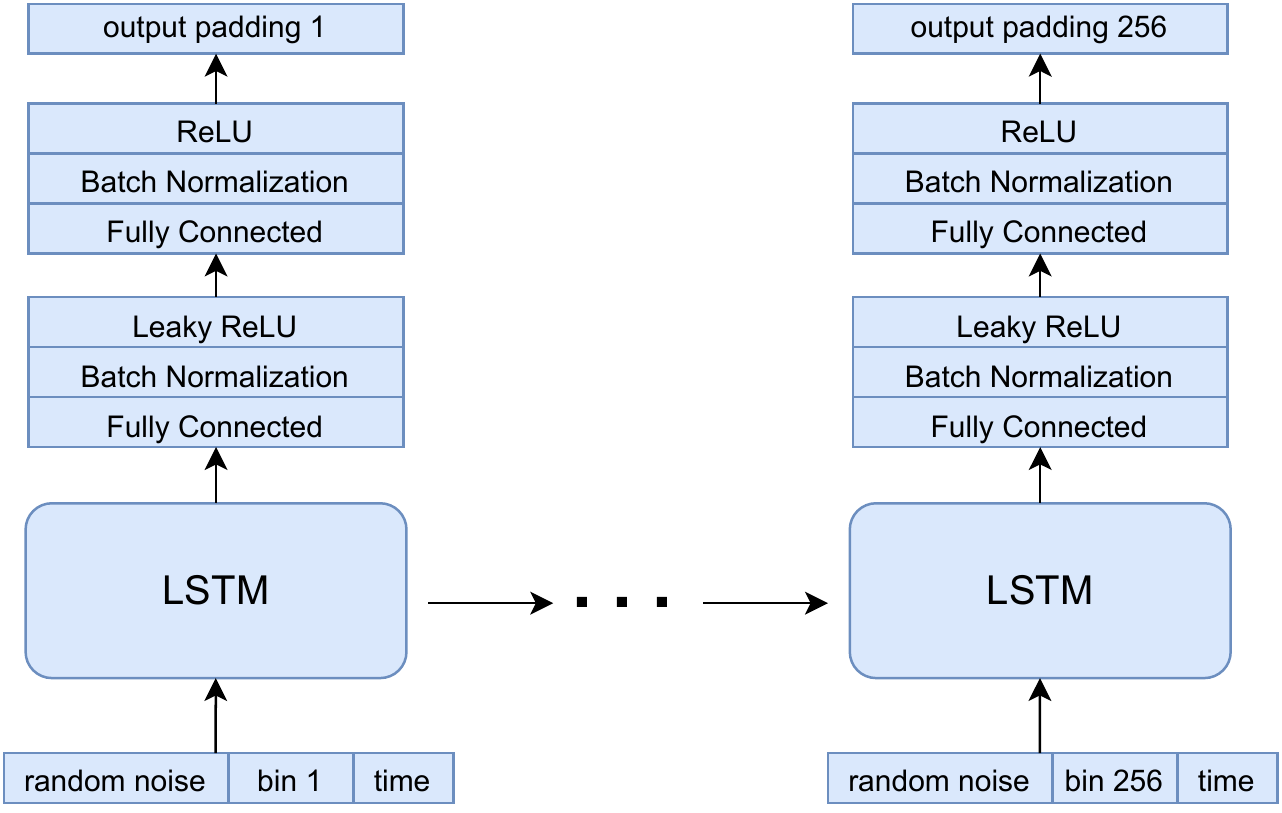}
  \caption{Unrolled LSTM generator outputting the dummy traffic volume at each step. Note that the volume in the previous bin, the time since the start of the defense, and random noise are used as LSTM input at each step.}
  \label{fig:lstm_unrolled}
\end{figure}

Considering that each bin is associated with a set time period, the LSTM can take as input random noise and the number of real packets in the associated \textit{previous} bin at each time step, then output the number of dummy packets that should be sent at the \textit{next} time step. This process is shown in Figure \ref{fig:lstm_unrolled}. While defense performance would be higher given the possibility to consider the number of real packets in the current bin of the trace representation, this cannot be known until the end of the associated time period, so the LSTM may only consider previous bins. Once bin 256 has been reached, padding stops until enough download traffic is sent to signal the loading of a new web page. Then, the defense restarts. The trace representation already ignores the first 9 packets of a page load, as described in Section \ref{trace_representation}, so we set this threshold to 9. Because the generator does not determine the amount of traffic in the first bin, this value is determined by randomly choosing a value from a uniform distribution ranging from 0 to 10. 

\subsection{Packet Timing}
\label{ssec:packet_timing}

\subsubsection{Download padding}
While the generator output specifies how many dummy packets to send in each interval, DeTorrent must still specify the exact timing. Web browsing traffic is often burst-heavy with occasional long gaps between bursts \cite{burstyshort, selfsimilar}, so we choose inter-packet delays from the exponential distribution, as done in some previous works \cite{compressive, watermarking}. The rate, $\lambda$, of the distribution is chosen such that the dummy traffic will on average fill the set bin (e.g. $\lambda$ is set such that the average inter-packet delay will be 1/10th of a second given 10 dummy packets and a bin length of one second). The beginning of the burst is randomly offset according to a uniform distribution within the bin to prevent an attacker from identifying dummy traffic based on bin timing. 

\subsubsection{Upload padding}

As mentioned earlier, Tor upload traffic tends to be distributed similarly in terms of when traffic bursts are sent, but at much lower volumes. As a result, DeTorrent's strategy is to simply add dummy traffic to maintain a set average of download-upload traffic. This achieves two goals: obfuscating much of the upload traffic and preventing the specific interleaving of the upload and download packets from leaking as much information. To be specific, DeTorrent sends dummy upload packets such that there is, on average, one upload packet for every five download packets. We chose this value for the parameter because, across all of the datasets used in this paper, few flows had a higher ratio of upload to download packets. This allows for DeTorrent to make nearly all the defended traces have the same download-packet ratio. To determine how many upload packets need to be added to achieve a given download-upload ratio, we must be able to create frequently updated estimates of the rate at which packets are sent or received. However, this is challenging due to the irregularity of internet traffic and the need for the method to be computationally inexpensive. As a result, we implemented a traffic volume estimator that weights recently sent traffic more heavily, making the estimate highly responsive to traffic bursts. For a more in-depth explanation of our method of estimating traffic rate, see the appendix Section \ref{ssec:timing_details}.

Using the frequently updated estimates, DeTorrent sends an upload packet whenever the upload traffic rate falls below one-fifth of the download traffic rate. If the upload traffic rate is larger than one-fifth of the download traffic rate, then no dummy traffic is sent. 
\section{DeTorrent Training}
\begin{figure*}
  \centering
  \begin{minipage}{0.48\textwidth}
    \centering
    \includegraphics[width=0.95\linewidth]{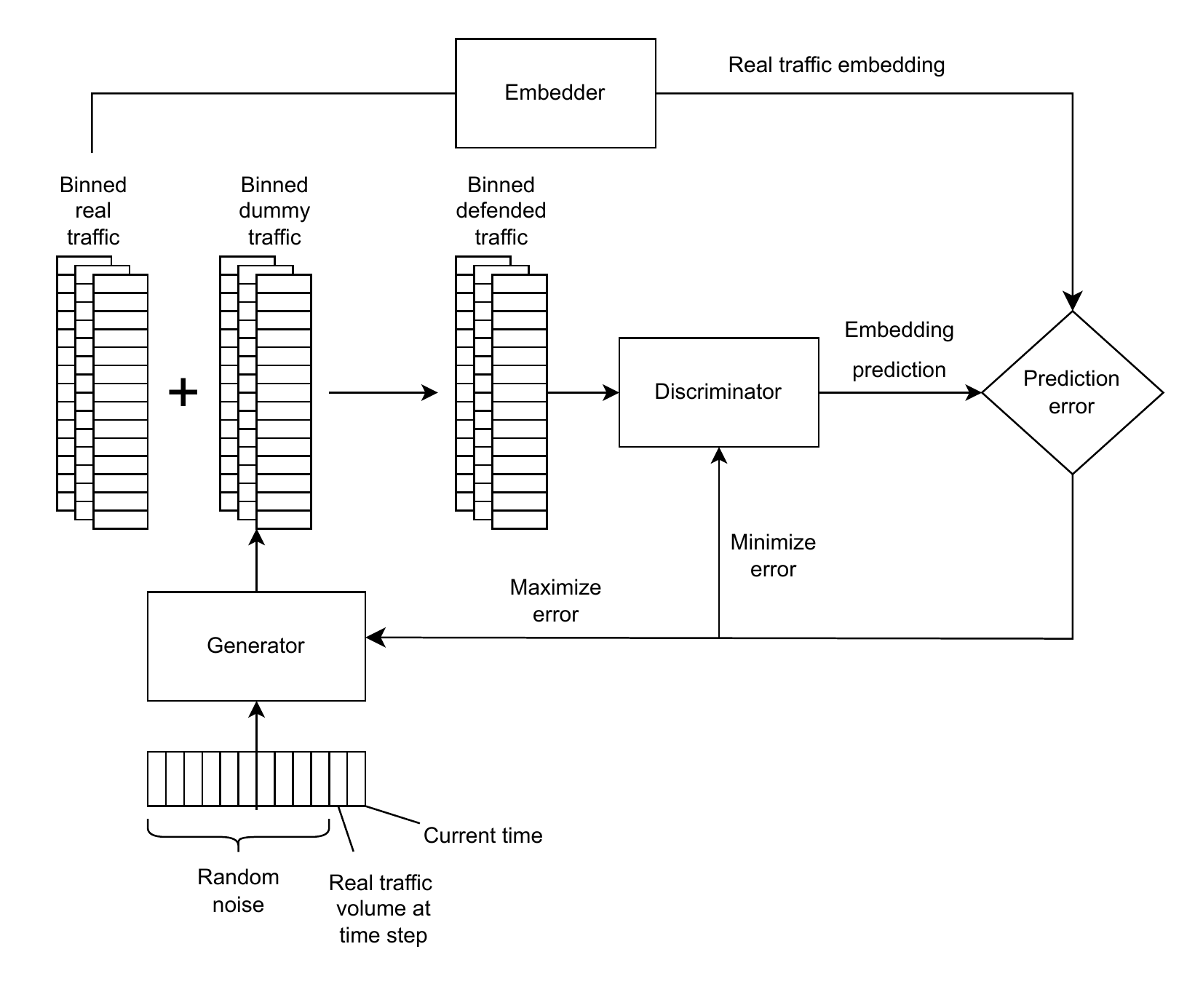}
    \caption{WF-DeTorrent training, where the generator minimizes the discriminator's ability to provide accurate embedding predictions. The binned real traffic and the binned dummy traffic are added to one another to create the defended trace, and the embeddings are computed using a pre-trained embedder. }
    \label{fig:wf_training}
  \end{minipage}
  \hfill
  \begin{minipage}{0.48\textwidth}
    \centering
    \includegraphics[width=0.95\linewidth]{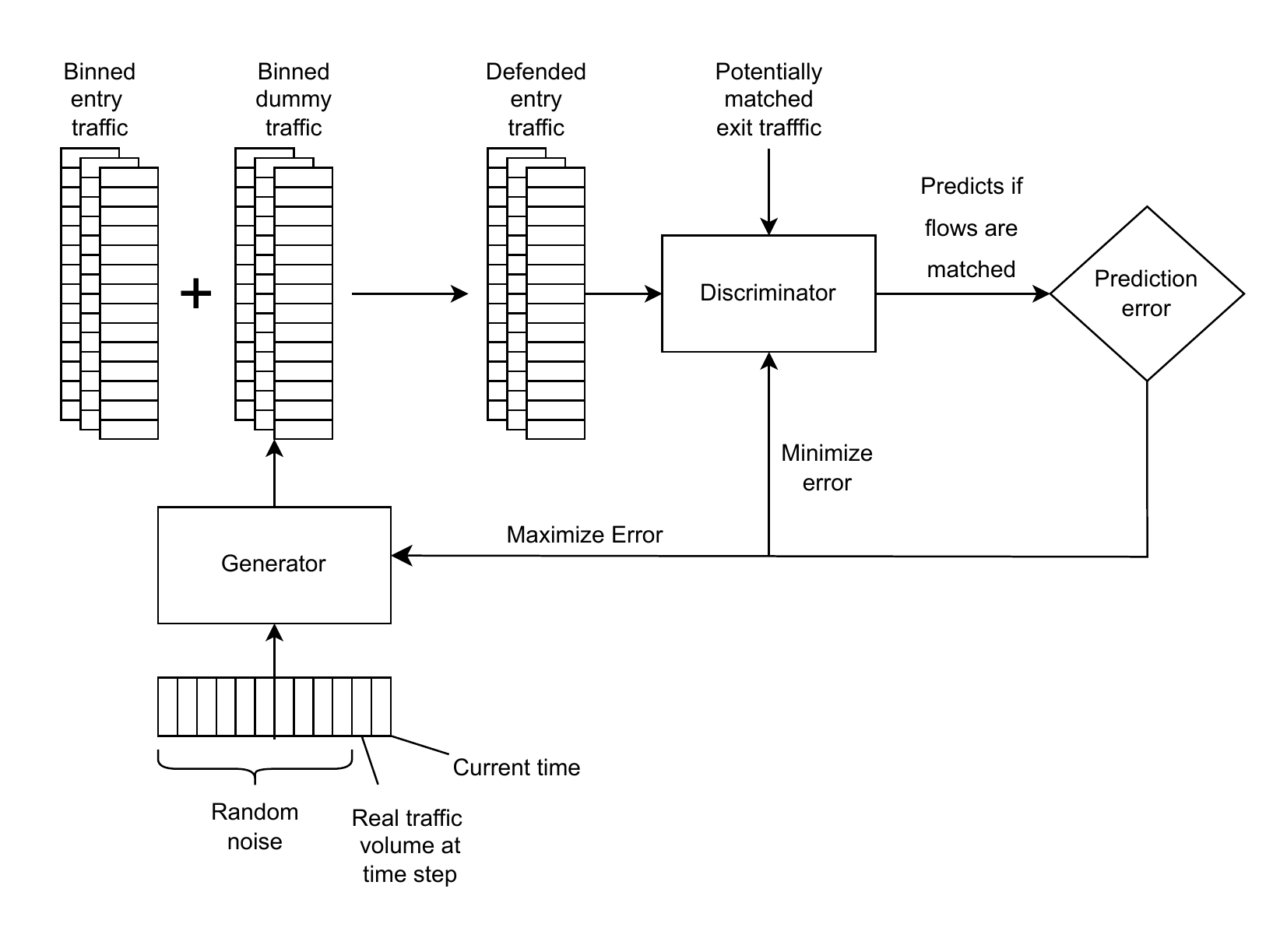}
    \caption{FC-DeTorrent Training, where the generator is trained to reduce the discriminator's ability to predict which flows are matched. The discriminator is trained to determine which flow pairs are associated with the same connection, and the generator is trained to defend the traffic to prevent this.}
    \label{fig:fc_training}
  \end{minipage}
\end{figure*}

\subsection{WF-DeTorrent Training}

In the WF setting, the discriminator and embedder are both initialized as convolutional neural networks that take a trace representation as input and output an embedding. The training process, described in Algorithm \ref{alg:wf_alg} and illustrated in Figure \ref{fig:wf_training}, starts by training the embedder using the previously described triplet loss method with Euclidean distance. Then, after a random trace is sampled, the generator is fed a tensor containing random noise and the number of real packets sent at that time step. Next, the generator outputs the proportion of the padding budget, $N_{download}$, to be sent at that time step. Once the final dummy padding volumes are calculated, they are added to the original trace to create a defended trace. Using the defended trace as input, the discriminator outputs a prediction of the embedding of the \textit{original} trace, indicating that it can accurately identify it. The discriminator's loss is computed as the distance between its prediction and the true embedding of the original trace. So, given original trace $x$ and noise $z$: \\

$l_d(x, z) = -l_g(x, z) = || D(x + G(z || x)) - E(x)||$\\

Note that the generator only uses the traffic volume information of the current bin at each time step, and that `+' indicates a bin-wise addition of the traces. Also, we can use the $N_{download}$ value to tune the amount of overhead DeTorrent uses. In the training loop, once the loss is computed and the discriminator's weights are updated, the noise is re-sampled and the same trace is defended again by the generator. This time the process is similar, but the generator's loss is computed as the negative of the discriminator's loss. Re-sampling the noise and running the generator again appears to aid convergence, likely because it discourages the generator and discriminator from overfitting on the most recent set of noise vectors. We also experimented with training the discriminator or generator for multiple batches at a time but found that training each for one batch led to the highest defense performance. 

We use the standard Gaussian to generate noise both because it is conventional with GANs and because the central limit theorem states that the sum of a large number of independent and identically distributed random variables is approximately normally distributed, regardless of the original distribution of the variables. Given the apparent universality to the Gaussian distribution, we hesitantly state that it is a reasonable default choice. 

Note that the goal of the generator is to pad the traffic so that the discriminator is unable to provide an accurate embedding. While it might seem more natural to have the generator try to thwart a WF classifier, consider that accurately embedding a trace is equivalent to classification in the sense that this embedding could be used as input for a simple classifier. Accordingly, we find that this setup is an effective proxy for defending traffic to resist classification by a WF adversary and solves the traffic representation problem discussed in Section \ref{ssec:trace_embedding}.

\begin{algorithm}
\caption{WF-DeTorrent Training}
\begin{algorithmic}

\STATE $E \leftarrow$ embedder model
\STATE $G \leftarrow$ generator model
\STATE $D \leftarrow$ discriminator model
\STATE $D_{train} \leftarrow$ training data
\STATE $\mathcal{L}_e \leftarrow$ embedder loss
\STATE $\mathcal{L}_g \leftarrow$ generator loss
\STATE $\mathcal{L}_d \leftarrow$ discriminator loss
\STATE $N_e \leftarrow$ number of epochs to train the embedder
\STATE $N_{adv} \leftarrow$ number of epochs to train the generator and discriminator
\STATE $m \leftarrow$ size of the generator input vector

\STATE

\FORALL{epoch in $N_e$}
\FORALL{batch $b_i$ in $D_{train}$}
\STATE sample anchors, positives from selected classes
\STATE sample negatives from different classes
\STATE anchor\_emb $\leftarrow E($anchor$)$
\STATE positive\_emb $\leftarrow E($positive$)$
\STATE negative\_emb $\leftarrow E($negative$)$
\STATE $\mathcal{L}_e \leftarrow \frac{1}{|b_i|}\sum_{A,P,N \in b_i} max(||E(A) - E(P) ||^2 - ||E(A) - E(N)||^2 + \alpha, 0)$
\STATE update $E(x)$ weights to minimize $\mathcal{L}_e$
\ENDFOR
\ENDFOR

\STATE
\FORALL{epoch in $N_{adv}$}
\FORALL{batch $b_i$ in $D_{train}$}
\STATE sample trace from $D_{train}$
\STATE sample noise $z_0,...,z_{m-1}$ from $\mathcal{N}(0,1)$
\STATE $\mathcal{L}_d = \frac{1}{|b_i|} \sum_{x \in b_i} || D(x + G(z_0,...,z_{m-1} || x)) - E(x)||$
\STATE update $D$ to minimize $\mathcal{L}_d$

\STATE 
\STATE sample noise $z_0,...,z_{m-1}$ from $\mathcal{N}(0,1)$
\STATE $\mathcal{L}_g \leftarrow -\frac{1}{|b_i|} \sum_{x \in b_i} || D(x + G(z_0,...,z_{m-1} || x)) - E(x)||$
\STATE update $G$ to minimize $\mathcal{L}_g$
\ENDFOR
\ENDFOR

\end{algorithmic}
\label{alg:wf_alg}
\end{algorithm}


\subsection{FC-DeTorrent Training}


The setup for FC-DeTorrent is similar in the sense that it consists of an LSTM generator `defending' traffic, which is then used as input to a CNN discriminator, as shown in Figure \ref{fig:fc_training}. However, here there is no embedder, as the generator directly minimizes the discriminator's ability to determine whether a given pair of flows is matched.

As described in Algorithm \ref{alg:fc_alg}, the training process starts by sampling entry and exit pairs. 50\% of the pairs are matched, as they may represent the entry and exit flows for the same traffic. Otherwise, they are unmatched. Then noise is sampled randomly and fed into the generator along with the number of real packets at each time step for the \textit{entry} flow. Once the generator has output the proportion of the dummy download traffic to send at each time step and multiplied it by $N_{download}$, it is added to the original entry flow traffic. This traffic is used as input to the discriminator, which tries to predict whether the defended entry and original exit are matched. Note that the entry is defended because dummy traffic can only be sent between the client and a relay in the Tor network. 

After making a batch of predictions, the discriminator's binary cross entropy loss is computed and its weights are updated. For entry flow e, exit flow x, true label y, and noise $z$: \\

$l_d(e, x, y, z) = y \log D(e + G(z||e), x) + (1-y)\log(1-D(e + G(z||e), x))$
\\

Next, the noise is re-sampled and the generator again outputs a padding defense which is added to the real traffic to create a defended trace. The discriminator then takes the defended trace and original exit trace and predicts if they are matched. This time, however, the weights of the generator are updated to maximize the discriminator's loss, thus incentivizing strategies that make it more difficult to correlate the matched entry and exit pairs. Like in the WF setting, the generator only has access to the current traffic volume at each step, and the generator's output is scaled to achieve a specific bandwidth overhead. 

\begin{algorithm}
\caption{FC-DeTorrent Training}
\begin{algorithmic}

\STATE $G \leftarrow$ generator model
\STATE $D \leftarrow$ discriminator model
\STATE $D_{train} \leftarrow$ training data
\STATE $\mathcal{L}_g \leftarrow$ generator loss
\STATE $\mathcal{L}_d \leftarrow$ discriminator loss
\STATE $N_{adv} \leftarrow$ number of epochs to train the generator and discriminator
\STATE $m \leftarrow$ size of the generator input vector
\STATE

\FORALL{epoch in $N_{adv}$}
\FORALL{batch $b_i$ in $D_{train}$}

\STATE sample $entry, exit$ pairs from $D_{train}$
\STATE sample noise $z_0,...,z_{m-1}$ from $\mathcal{N}(0,1)$
\STATE $\mathcal{L}_d = \frac{1}{|b_i|} \sum_{entry, exit \in D_{train}} l_d(entry, exit, z_0,...,z_{m-1})$
\STATE Update $D$ to minimize $\mathcal{L}_d$

\STATE 
\STATE sample noise $z_0,...,z_{m-1}$ from $\mathcal{N}(0,1)$
\STATE $\mathcal{L}_g = -\frac{1}{|b_i|} \sum_{entry, exit \in D_{train}} l_d(entry, exit, z_0,...,z_{m-1})$
\STATE update $G$ to minimize $\mathcal{L}_g$
\ENDFOR
\ENDFOR

\end{algorithmic}
\label{alg:fc_alg}
\end{algorithm}

\section{Experiment Details}
\subsection{Datasets}

The BigEnough dataset, referred to here as `BE,' was collected for a Systemization of Knowledge paper on website fingerprinting defenses \cite{evaluation} by Matthews et al. The BE collection methodology was based on that of Pulls for the GoodEnough dataset \cite{machines} and contains three modes based on the Tor Browser security configurations: Standard, Safer, and Safest, though we only use the default standard mode in this paper. Most importantly, the BE dataset considers 95 websites represented by 10 subpages which are each crawled 20 times, resulting in 19,000 traces. Including subpages increases the realism of the dataset compared to past datasets, which model traffic as always visiting the home page of a given website. The Open PageRank Initiative \cite{pagerank} top website list was used to select the most popular websites. The BigEnough dataset also includes Tor log information needed to simulate the Spring and Interspace defenses for comparison; unlike other datasets, it also records the timing of Tor cells rather than that of the associated packets. 

We use the website fingerprinting dataset by Sirinam et al., originally collected to demonstrate the Deep Fingerprinting attack \cite{Sirinam2018}, to evaluate defense performance against an attacker able to compile a large and robust dataset. The dataset was collected by visiting the home pages of the Alexa top 100 websites 1,250 times each using ten local machines. The crawl was done in batches to account for both long and short-term variance following the methodology described by Wang et al. \cite{Wang2013}. After filtering, we are left with a dataset with 1000 instances of 95 websites, which we refer to as `DF.'  

To test the flow correlation defense performance, we use the dataset collected by Oh et al., referred to here as `DCF,' to demonstrate the effectiveness of the DeepCoFFEA attack \cite{dcf}. Their dataset collection methodology was based on that of the DeepCorr technique by Nasr et al. \cite{deepcorr}, though with some changes. Specifically, they used a web crawler along with a physical machine to collect the entry flows and a proxy server to collect the exit flows. Over 60,000 Alexa websites were crawled in over 1,000 batches with the first 60 seconds of each flow being recorded. The DCF dataset removes flows with less than 70 packets total or with a window packet count of less than 10. While the original dataset includes 42,489 flow pairs, we randomly selected 20,000 in our experiments. 

To evaluate the DeTorrent pluggable transport defense, we visited each website in the Alexa top 250 \cite{alexa} 100 times over two weeks in September 2022. To collect the dataset, we used a fork of Tor Browser Crawler \cite{crawler} updated to work with more recent libraries and Tor Browser version 11.0.15. The crawler then uses the DeTorrent pluggable transport, described in Section \ref{ssec:real_implementation}, to transform traffic between it and the bridge. We ran two separate instances of the crawler, each in its own virtual machine, and connected to one of two Tor bridges. We hosted the bridges using Amazon Lightsail and used Tor version 0.4.7.8. In this paper, we refer to this dataset as `PT.'

\subsection{WF Attack Details}
\label{ssec:wf_details}
We used open-source code associated with WTF-PAD \cite{wtfpadcode}, Interspace \cite{pullscode}, Spring \cite{pullscode}, FRONT \cite{frontcode}, and BAP \cite{bapcode} to generate the defended datasets for each defense. Then, to test the robustness of WF-DeTorrent and comparable defenses, we test the defense's ability to resist state-of-the-art attacks. For the website fingerprinting setting, we use the Tik-Tok and Deep Fingerprinting attacks. These attacks are based on the same CNN architecture, though Tik-Tok uses packet timing information, allowing it to achieve somewhat higher accuracy. We use the default parameters with the exception that we run the Tik-Tok and Deep Fingerprinting attacks for 100 epochs with a patience value of 10. This ensures that training does not stop early on the BE dataset, which is much smaller than the DF dataset and thus has much smaller epochs. We also increased the attack input size from 5,000 to 10,000 to account for the increased traffic volume observed in the BE, DCF, and PT datasets (compared to the DF dataset). For both Tik-Tok and Deep Fingerprinting, we use 5-fold cross-validation to estimate attack performance. 

To fairly compare WF-DeTorrent to defenses such as FRONT and WTF-PAD, we scale the bandwidth overhead to match the overhead used by WF-DeTorrent. To scale FRONT, we increase the upload and download volume parameters $N_s$ and $N_c$. To scale WTF-PAD using the distributions released in the original paper, we scale the timestamps of the datasets, use the WTF-PAD simulator to insert the dummy padding, and then re-scale the timestamps to match the original trace lengths. This strategy is used in past literature \cite{machines} and is necessary due to the smaller inter-packet delays seen in more recently collected datasets. To simulate BAP, we train the timing and direction perturbation attacks against Var-CNN as indicated by the source code shared by the authors. We also set the $\alpha$ parameter to tune the volume of the perturbations. The specific overheads and defense parameters are shown in Table \ref{tab:df_overhead} and Table \ref{tab:be_overhead}. Also, it should be noted that a low but measurable amount of added latency may be incurred by a full deployment of these `zero-delay' defenses, as shown in past work \cite{paddingonly}. However, the exact amount of additional latency is difficult to estimate without a larger-scale Tor simulation. 

The default Spring and Interspace defenses use a comparable amount of bandwidth overhead. Because the output of the circuit padding simulator \cite{pullscode} contains unrealistically precise timing information, we round the timestamps of the defended dataset to the nearest hundredth of a second. We only simulate Spring and Interspace on the BigEnough dataset, as their simulator requires Tor logs that are not present in the DF dataset. 

To simulate WF-DeTorrent on a dataset, we must take steps to prevent it from being able to defend traces that it witnesses during training: this might allow it to memorize effective defense strategies and perform unrealistically well. So, we partition each dataset $n$-ways, where $n-2$ partitions are used for training, one is used for validation, and the other is used to simulate WF-DeTorrent and output the defended traces. Then, we rotate which partitions are used for training, evaluation, and output so that each partition is a test set exactly once, creating the defended dataset. While we do scale $N_{download}$ based on the choice of dataset, we do not do extensive hyperparameter optimization to tune DeTorrent, as it takes several hours to train in this manner. 

To make clear comparisons between the website fingerprinting defenses, we test them in the closed-world setting where the attacker simply needs to determine which website a trace is associated with. While this setting is less realistic than the open-world setting, we prefer to test the defenses in a setting more advantageous to the attacker, as we can be certain that defense performance is not likely due to the experimental design. Furthermore, we make assumptions about the adversary's ability. These include that the adversary can determine the beginning and end of the page load, that the user visits the home page of each website, and that the adversary can create a dataset that accurately represents the target's unique network conditions. Again, these assumptions favor the attacker, so we expect real-world website fingerprinting to be at least as difficult as our results would indicate.

\subsection{FC Attack Details}

In the flow correlation setting, we use the DCF dataset and the DeepCoFFEA attack with default parameters to test the defenses. 10,000 flows from the dataset are used for training and 10,000 are used for testing. We train DeepCoFFEA for 1,500 epochs, as this tends to lead to the highest attack performance. 

To simulate BAP in the FC setting, we train it to generate perturbations using the DeepCorr attack provided in the open-source BAP implementation \cite{banp}. We also set the $\alpha$ parameter to determine the volume of the traffic perturbations. For FRONT, we set the $N_s$ and $N_c$ parameters to achieve a comparable level of bandwidth overhead. To set the WTF-PAD overhead, we scale the timestamps of the DCF dataset, use the normal\_rcv distributions as specified in the original paper, and then re-scale the timestamps to match the original. The specific parameter and bandwidth overhead value for the defenses are shown in Table \ref{tab:dcf_overhead}. We don't use the Spring or Interspace defenses in the FC setting, as they require Tor log information that the DCF dataset does not contain. To implement the Decaf defense, we randomly selected windows of traffic from dummy traces in the DeepCorr dataset \cite{deepcorr} and used them to pad the real traffic traces, as described in the original `Decaf-DC' implementation. To roughly match the volume of dummy traffic used by other defenses, we choose to apply the defense to all of the real traffic windows, rather than randomly selecting them. 

To prevent FC-DeTorrent from having the unfair advantage of being able to train the generator and discriminator on the same dataset that it defends, we use the dataset partitioning method described in Section \ref{ssec:wf_details}. We again refrain from doing hyperparameter optimization to tune FC-DeTorrent's training and implementation.

\subsection{Real-World Implementation Details}
\label{ssec:real_implementation}
To demonstrate the practicality of the DeTorrent defense, we implement it as a pluggable transport \cite{pt}. Pluggable transports (PTs) transform traffic between the client and a Tor bridge \cite{bridge}, which is a Tor relay that is not publicly listed. PTs are frequently used for censorship circumvention due to their ability to obfuscate traffic characteristics. We implemented DeTorrent on top of the WFPadTools framework \cite{wfpadtools}, which extends the Obfsproxy PT \cite{obfsproxy} while adding functionality for common anti-traffic analysis strategies. The DeTorrent PT uses a private Tor bridge, which we run using Amazon Lightsail. Even though DeTorrent requires the LSTM generator to be repeatedly evaluated, we are able to defend five connections simultaneously with a modest virtual private server with 4GB RAM and 2 virtual CPUs running at less than 50\% capacity. Thus, DeTorrent is not too computationally intensive for real-time use and does not require a GPU for implementation.

\section{Website Fingerprinting Results}

\begin{table*}
    \small
    \centering

    \scalebox{.9}{
    \begin{tabular}{p{2cm}||p{2cm}|p{7cm}}

    \hline
    Defenses & Bandwidth OH & Parameters\\
    \hline
    WTF-PAD & 86.1\% & normal\_rcv, scaled 2x\\
    FRONT & 90.2\% & $N_s = N_c = 1700, W_{min}=1, W_{max}=14$\\
    BAP & 90.2\% & $\alpha=2000, \sigma=0, \mu=0$\\
    WF-DeTorrent & 88.8\% & $N_{download}=1500$\\
    \hline

\end{tabular}}
    \vspace{5pt}
    \captionsetup{justification=centering,margin=2cm}
    \caption{Defense Overheads and Parameters on DF}
    \label{tab:df_overhead}
\end{table*}

\begin{table*}[t]
    \small
    \centering
    \scalebox{.9}{
    \begin{tabular}{p{2cm}||p{2cm}|p{7cm}}

    \hline
    Defenses & Bandwidth OH & Parameters\\
    \hline
    WTF-PAD & 86.8\% & normal\_rcv, scaled 6x\\
    BAP & 94.4\% & $\alpha = 5000, \sigma=0, \mu=0$\\
    Spring & 102.7\% & see \cite{machines}\\
    Interspace & 96.7\% & see \cite{machines}\\
    FRONT & 95.0\% & $N_s = N_c = 4000, W_{min}=1, W_{max}=14$\\
    WF-DeTorrent & 98.9\% & $N_{download}=3000$\\
    \hline

\end{tabular}}
    \vspace{5pt}
    \captionsetup{justification=centering,margin=2cm}
    \caption{Defense Overheads and Parameters on BE}
    \label{tab:be_overhead}
\end{table*}

\begin{table*}[t]
    \centering
    \begin{minipage}{0.45\textwidth}
        \centering
        \scalebox{.9}{
        \begin{tabular}{p{2.5cm}||p{2cm}|p{2cm}}
            \hline
            Defenses & Tik-Tok & DF\\
            \hline
            \hline
            Undefended & 97.7\% & 98.1\%\\
            BAP & 94.4\% & 86.8\%\\
            WTF-PAD & 92.1\% & 90.3\%\\
            FRONT & 85.2\% & 79.7\%\\
            WF-DeTorrent & 79.5\% & 74.5\%\\
            \hline
        \end{tabular}}
        \caption{Closed-World Accuracy on DF}
        \label{tab:df_accuracy}
    \end{minipage}\hfill
    \begin{minipage}{0.45\textwidth}
        \centering
        \scalebox{.9}{
        \begin{tabular}{p{2.5cm}||p{2cm}|p{2cm}}
            \hline
            Defenses & Tik-Tok & DF\\
            \hline
            \hline
            Undefended & 93.4\% & 94.3\%\\
            BAP & 92.4\% & 94.3\%\\
            WTF-PAD & 62.4\% & 59.7\%\\
            Spring & 51.4\% & 46.8\%\\
            Interspace & 44.9\% & 38.1\%\\
            FRONT & 42.4\% & 41.2\%\\
            WF-DeTorrent & 31.9\% & 30.0\%\\
            \hline
        \end{tabular}}
        \caption{Closed-World Accuracy on BE}
        \label{tab:be_accuracy}
    \end{minipage}
\end{table*}

\begin{table*}[htbp]
    \small
    \centering
    \scalebox{.9}{
    \begin{tabular}{p{2cm}||p{2cm}|p{7cm}}

    \hline
    Defenses & Bandwidth OH & Parameters\\
    \hline
    WTF-PAD & 98.6\% & normal\_rcv, scaled 3.2x\\
    BAP & 96.6\% & $\alpha = 4000, \sigma=0, \mu=0$\\
    FRONT & 97.9\% & $N_s = N_c = 3400, W_{min}=1, W_{max}=14$\\
    Decaf & 79.6\% & $\omega = 5, \frac{v}{k}=1$\\
    FC-DeTorrent & 97.3\% & $N_{download}=3200$\\
    \hline

\end{tabular}}
    \vspace{5pt}
    \captionsetup{justification=centering,margin=2cm}
    \caption{Defense Overheads on DCF}
    \label{tab:dcf_overhead}
\end{table*}

\subsection{Simulated Defense Results}

Defense performance on the BE dataset is shown in Table \ref{tab:be_accuracy}. On the BE dataset, we find that WF-DeTorrent reduces both the Tik-Tok and DF attacks to the lowest accuracy at 31.9\% and 30.0\% respectively. Because WF-DeTorrent is trained to `trick' the discriminator into outputting inaccurate embeddings of the target traffic by making traffic from different web pages indistinguishable, we believe it is able to effectively mask the differences between many of the subpages. Thus, the WF attacks are usually unable to uniquely identify a given subpage in the BE dataset. 

FRONT is the next-best performing defense with Tik-Tok and DF accuracies of 42.4\% and 41.2\% respectively. FRONT likely performs well because the inherent randomness in the shape and volume of the dummy packet distribution makes training the WF attacks more difficult. This effect appears exacerbated by the fact that the BE dataset is relatively small and distributes the instances for each website across ten subpages.

WTF-PAD provides a moderate amount of defense by masking the large inter-packet delays, which allows for the traces to be more easily identified. However, WTF-PAD defends in a fairly consistent manner without much obfuscation of the volume of quick bursts of traffic. Thus, many of the traces can still be correctly classified. 

The Spring defense outperforms WTF-PAD, showing that an `evolved' circuit padding defense can be somewhat effective. However, the Spring defense strategy is relatively consistent across traces, allowing for a DNN to effectively learn how to classify the defended traces. The Interspace defense, on the other hand, is `probabilistically defined,' in the sense that it is initialized with one of a set of possible padding strategies at the client and relay. Interspace also randomizes the parameters for the length and inter-arrival time distributions. As a result, Interspace is significantly more effective than its Spring counterpart. BAP, however, provides little defense against the Tik-Tok and DF attacks. This is likely because the BAP perturbation generator is trained to reduce the accuracy of a \textit{static} attacker that cannot retrain. In this setting, the WF attacks can be retrained so that they are mostly unaffected by the BAP defense. 

Defense results on the DF dataset are shown in Table \ref{tab:df_accuracy}. In terms of performance, the order of defenses is the same except that we are unable to evaluate Spring and Interspace, which require Tor logs to simulate. In general, the defenses did not reduce the accuracy nearly as much. However, this is likely due to the inherent difficulty of defending the DF dataset, which has 1,000 instances for each class, allowing for the deep learning-based WF attacks to better generalize. The DF dataset also uses only the home pages of the websites, meaning there is less intra-class variety. This especially reduces the effectiveness of defenses that utilize randomized dummy packet distributions, such as DeTorrent and FRONT. We believe that this is because this makes it much less likely that the defended trace for one class will be `confused' with the trace of another class.

\subsection{Trade-off Between Overhead and Performance}

\begin{figure}
  \centering
  \includegraphics[scale=.55]{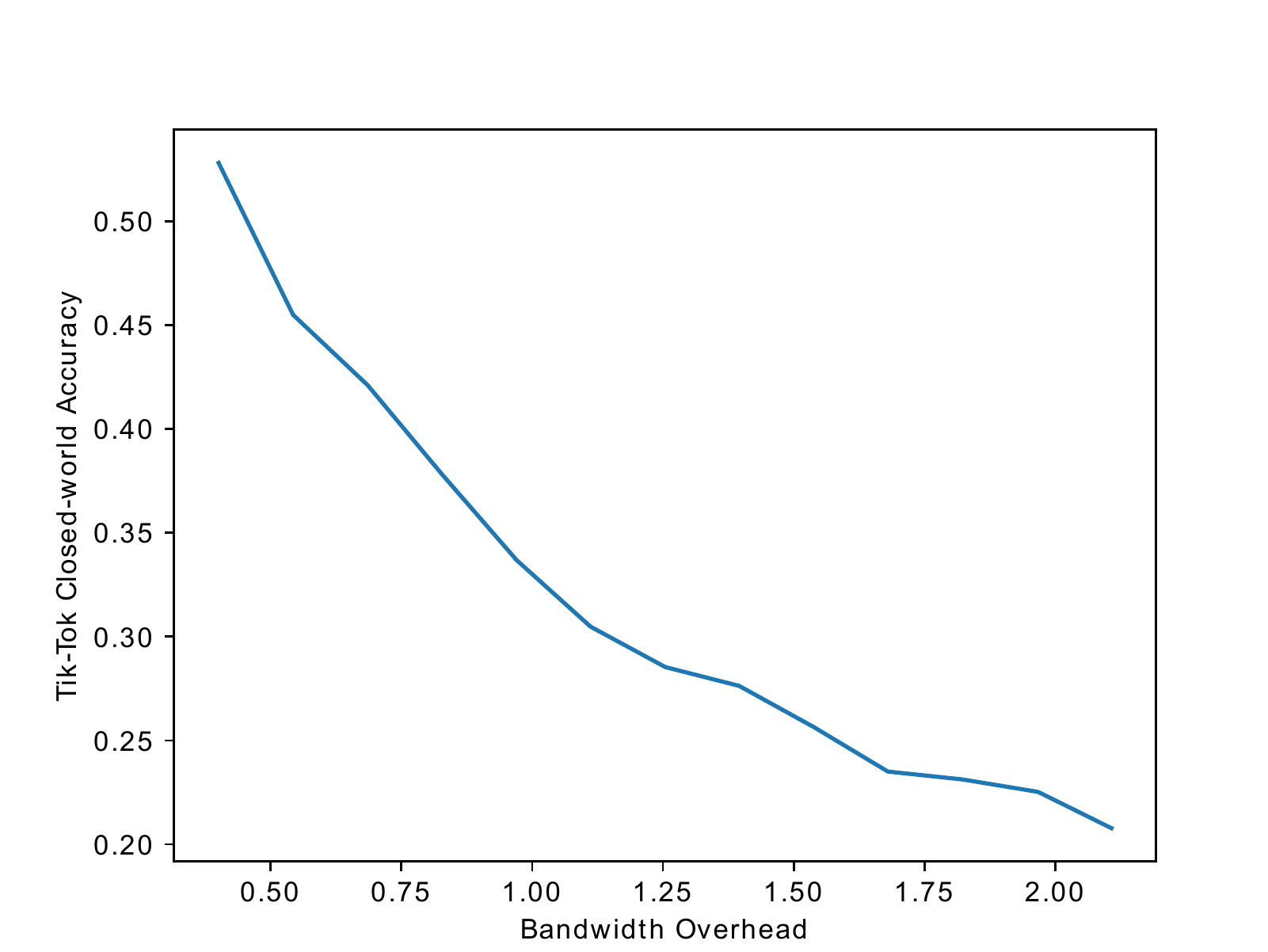}

  \caption{DeTorrent Overhead vs. Performance}
  \label{fig:tradeoff}
\end{figure}

In order to fully evaluate the relationship between DeTorrent's performance and bandwidth overhead, we tested its ability to defend traffic in the BE dataset against the Tik-Tok attack for a variety of download volume ($N_{download}$) parameter choices. We varied $N_{download}$ from 1,000 to 7,000 while keeping the upload traffic ratio constant, resulting in overhead that varied from about 40\% to about 210\%, as shown in Figure \ref{fig:tradeoff}. Tik-Tok attack accuracy was 52.8\% for the lowest bandwidth setting and was reduced to 20.8\%. We also find that increasing DeTorrent's bandwidth budget initially improves its performance, but is later subject to decreasing returns. To be specific, increasing $N_{download}$ from 1000 to 3000 decreases Tik-Tok performance by about 19.1\%, while increasing $N_{download}$ from 5000 to 7000 only further decreases it by 4.9\%. This is likely due to the difficulty of defending particularly unique and high-volume traces.

Additionally, for comparison to the next-best non-scaled WF defense, we simulate FRONT on the BE dataset while using the original, non-scaled parameters. We find that this version of FRONT incurs a 59.5\% bandwidth overhead while reducing Tik-Tok attack accuracy to 54.3\%. For comparison, WF-DeTorrent with parameter $N_{download} = 1500$ defends BE with an overhead of 54.3\% while reducing Tik-Tok attack accuracy to just 43.8\%.

\subsection{Trace Representation Tuning}

Because the choice of trace representation can significantly impact performance, we evaluate WF-DeTorrent's performance for different representation lengths and binning strategies. As shown in Figure \ref{subfig:representation}, we find that the performance increases for larger trace representations until about length 512, likely because the granularity of the representation allows for WF-DeTorrent to make more fine-tuned strategies. However, WF-DeTorrent performance begins to decrease for larger representation sizes. We believe that this is due to the difficulty of accurately embedding the traces when bins are very small, as slight timing differences begin to change traffic representation significantly. As a result, we choose to use a trace representation of length 256, as it is associated with high performance while being less computationally intensive to train and implement compared to larger representations. 

We also test a representation that spaces the bins evenly on a linear scale rather than on a logarithmic scale and find that this significantly decreases defense performance, reducing Tik-Tok to 43.2\% rather than 31.9\%. This decrease is likely because the linear scale does not as effectively differentiate between the frequent bursts of packets often seen early in a trace.

\subsection{Defense Countermeasures}

Since adversarial training allows the discriminator to adapt to the generator's strategy, we don't expect straightforward improvements to attacks based only on knowledge of the use of DeTorrent. Still, an attacker aware of the DeTorrent defense in use may use hyperparameter tuning to increase attack accuracy. During our WF defense evaluation, we modified the epoch number, input dimension, early stopping parameters, learning rate, and optimizer type. We find that the attack accuracy is highest when setting the number of epochs to 100, increasing the input dimension to 10,000, and setting the early stopping patience value to 10. Similarly, we modify the number of epochs used in the DCF attack to 1,500 and check validation accuracy to best improve performance. However, changing the other hyperparameters in the WF and FC settings did not significantly change performance.

However, one countermeasure that an attacker can use to better evade a defense is to generate multiple defended traces for every real trace in the dataset, as demonstrated in an evaluation of a variety of WF defenses by Matthews et al. \cite{evaluation}. This works particularly well when using small datasets and against defenses that use randomization, likely because it aids attacker model generalization and provides a better view of the variety of ways in which traffic may be defended. To test this, we generate a `10x' dataset where each trace in the training set is defended 10 times each (with re-sampled noise) to create a much larger training set. We find that this improves Tik-Tok attack performance from 31.9\% to 48.2\%. Thus, a motivated attacker would likely use this countermeasure to best attack DeTorrent (as well as comparable defenses such as FRONT). 

\subsection{Real-World Implementation Results}

Then, we crawl a set of popular websites through the pluggable transport to test current website fingerprinting defenses against the collected packet traces. To test the transferability of the defense, the DeTorrent download padding distribution is determined using a generator trained using the BE dataset and scaled to achieve a similar bandwidth overhead. As a result, the generator is not fine-tuned to current internet traffic. However, since DeTorrent's defense performance transfers well to different settings and sets of websites, we find that defense performance remains high, reducing Tik-Tok to a 21.5\% accuracy with an 88.8\% bandwidth overhead.

\begin{figure}
  \centering
  
  \begin{minipage}{\linewidth}
    \centering
    \includegraphics[scale=.55]{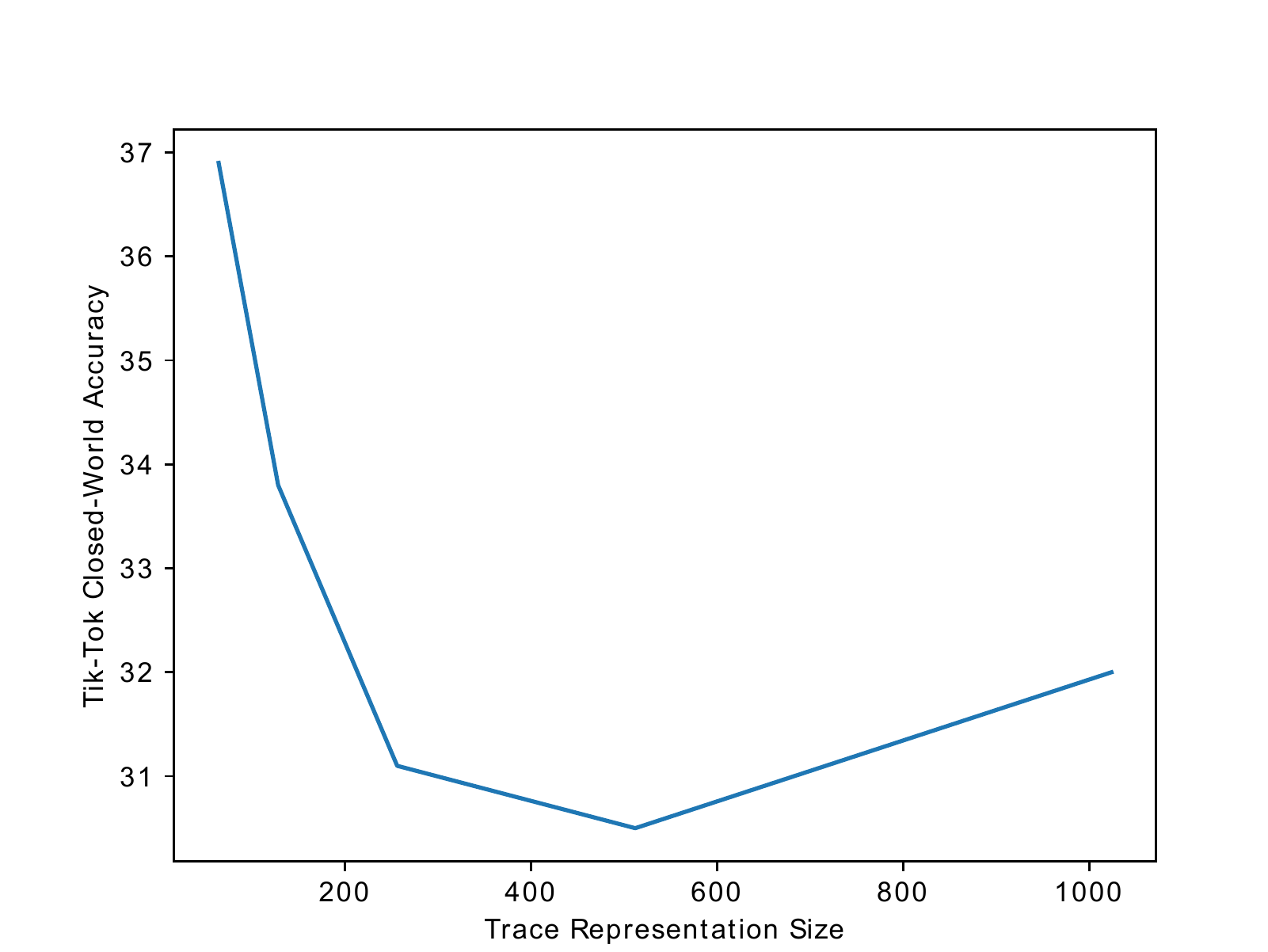}
    \par
    \caption{DeTorrent Performance vs. Representation Length}
    \label{subfig:representation}
  \end{minipage}
  
  \vspace{10pt} 
  
  \begin{minipage}{\linewidth}
    \centering
    \includegraphics[scale=.55]{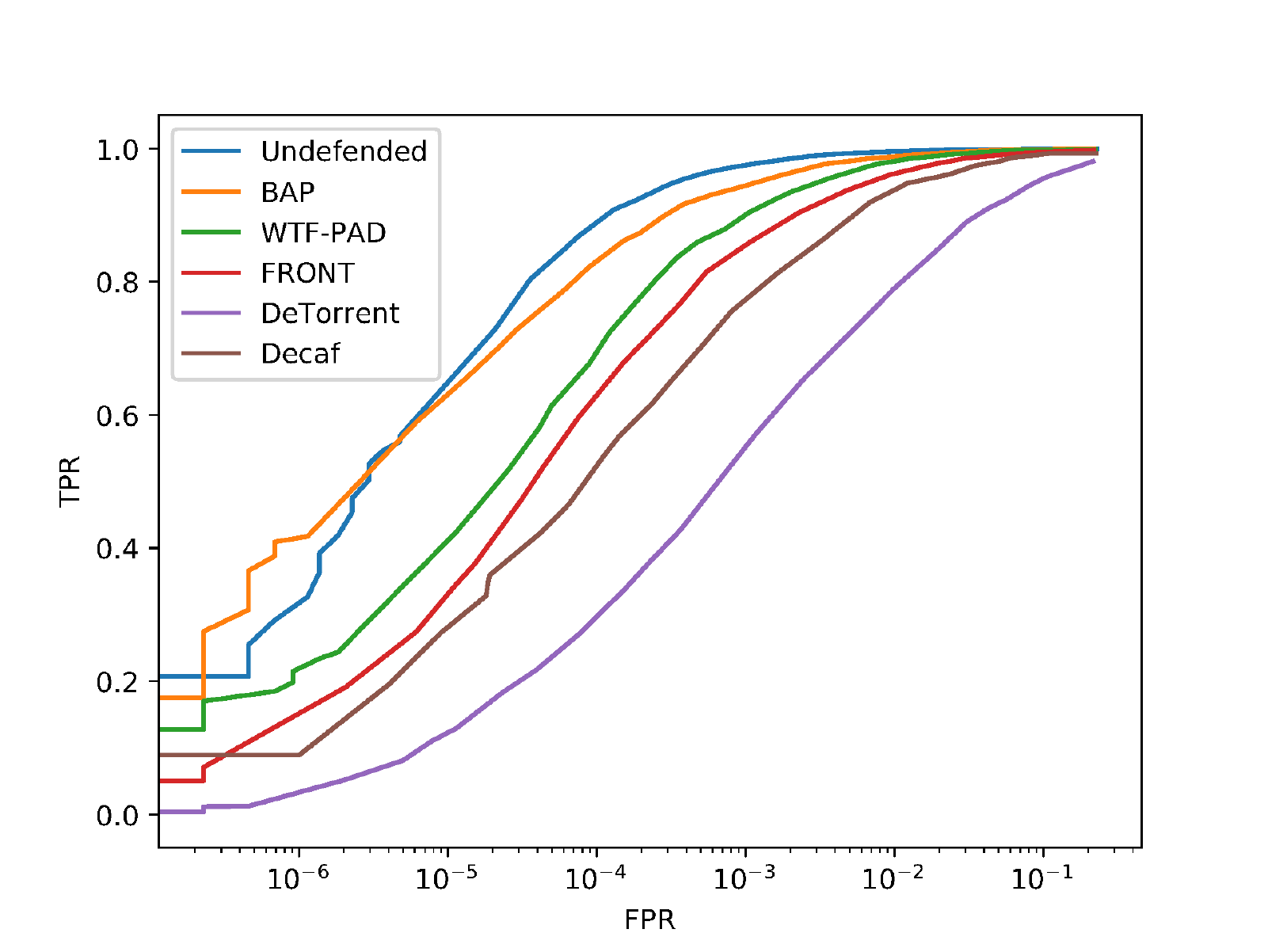}
    \par
    \caption{Flow correlation defense performance against DeepCoFFEA. Note that lower TPR implies a better defense.}
    \label{subfig:fc_performance}
  \end{minipage}
  
   \label{fig:combined}
\end{figure}

\section{Flow Correlation Results}

To ensure a fair comparison between the chosen defenses, we scale them to operate with comparable bandwidth overhead, as shown in Table \ref{tab:dcf_overhead}. Also, one of the main challenges of the flow correlation setting is achieving a high enough true positive rate without allowing the number of false positives to outnumber the true positives. As a result, it is most sensible to evaluate flow correlation defenses by comparing their true positive and false positive rates at various DeepCoFFEA correlation thresholds; this is illustrated in Figure \ref{subfig:fc_performance}.

In terms of reducing DeepCoFFEA's performance, FC-DeTorrent far outperforms the compared defenses, as shown in Figure \ref{subfig:fc_performance}. While other defenses allow a high true positive rate for associated false positive rates of as little as $10^{-3}$, FC-DeTorrent noticeably reduces DeepCoFFEA's true positive rate for all but the highest false positive rates. For a $10^{-4}$ false positive rate, FC-DeTorrent reduced  DeepCoFFEA's true positive rate to .30, compared to about .54 for Decaf and .63 for FRONT. For a $10^{-5}$ false positive rate, the true positive rates are about .13 for FC-DeTorrent, .30 for Decaf, .34 for FRONT, and .42 for WTF-PAD. FC-DeTorrent performance is likely due to its ability to learn how to pad entry flows such that they are easily confused with non-associated entry flows, thus confusing the discriminator. Also, FC-DeTorrent's defense advantage over the alternate FC defenses is significantly larger than WF-DeTorrent's advantage in the WF setting: we believe that this is because FC-DeTorrent can directly train against the discriminator, while WF-DeTorrent must rely on the embedder to estimate descriminator performance.

FRONT provides a reasonable degree of FC defense, likely because it's randomized and because it sends traffic according to the Rayleigh distribution, which mimics the distribution of real traffic. As a result, the defended entry traffic associated with a given exit flow is likely to look similar to other defended entry flows. The same can be said of Decaf, which defends the real traffic using traffic patterns found in previously collected traffic. However, FRONT's use of the Rayleigh distribution makes the padding somewhat predictable, as the amount of padding will always increase until it peaks and will then slowly taper off. As a result, DeepCoFFEA is able to learn how to effectively ignore much of the padding. Decaf, on the other hand, avoids this problem by using more realistic traffic patterns and therefore is a more performant defense. 

WTF-PAD obfuscated some of the inter-packet delays that DeepCoFFEA uses to identify traffic; as a result, it provides some defense as well. Still, WTF-PAD does not place much emphasis on the more coarse features of the traffic, such as the sizes of the largest traffic bursts. As a result, DeepCoFFEA is still able to correlate many of the flows defended by WTF-PAD. Because BAP was trained to provide adversarial perturbations against a static model, DeepCoFFEA's feature embedders can retrain against the BAP-defended dataset. As a result, BAP's performance is lower than that of other defenses.

\section{Discussion}
\subsection{Transferability}
\subsubsection{Across Different Datasets}

It is important to ensure that DeTorrent is able to effectively defend traffic with different characteristics than traffic in the training datasets. Otherwise, DeTorrent's performance may degrade rapidly over time or while defending new websites. Accordingly, we test DeTorrent's transferability by conducting an experiment where we train WF-DeTorrent on traces associated with one set of websites while testing the defense on traces associated with a \textit{different} set. 

More specifically, we partition the BE dataset so that the first partition contains 20 instances of 10 subpages of 50 of the websites, while the second partition contains the same number of instances of the other 45 websites. We refer to the first partition as BE-1 and the second partition as BE-2. Then, we train the WF-DeTorrent defense using only the BE-1 dataset and simulate the defense on BE-2 (which does not contain any traces \textit{or} websites that WF-DeTorrent was trained on) with $N_{download} = 3000$. We find that the defense drops Tik-Tok's accuracy to 40.3\%, showing that it was effective. Also, note that the higher attack accuracy is likely because the attacker is only choosing from 50 websites, rather than 95. 

To set a baseline for how well WF-DeTorrent performs when training and testing on the same dataset, we both train and test it on BE-2 with the same amount of overhead. We find that it decreases Tik-Tok's performance to 39.6\%. Because this accuracy is only .7\% lower than the accuracy found when WF-DeTorrent was trained on BE-1, we conclude that the WF-DeTorrent framework generalizes well and does not simply memorize fine-grained patterns in the training data. Accordingly, we expect the defense provided by WF-DeTorrent to be highly transferable.

\subsubsection{Against Different Attacks}

In this paper, we describe two different training frameworks with one defending against website fingerprinting and the other defending against flow correlation. However, one might like to defend against both types of attacks. To test how DeTorrent's performance against one attack transfers to the other, we test how well WF-DeTorrent functions in the FC setting and vice versa. To train the WF and FC DeTorrent models, we use the BE and DCF datasets respectively. Note that WF-DeTorrent and FC-DeTorrent are trained on entirely different datasets collected using different methodologies and at different points in time. As a result, the models used in these experiments are unlikely to be fine-tuned in terms of defense performance, and the above results should be taken as a lower bound of what performance is possible. 

We find that FC-DeTorrent provides some defense in the WF setting, though its performance is degraded. To be specific, it reduces Tik-Tok performance on the BE dataset to 45.8\%. This represents a 12.9\% increase in Tik-Tok accuracy, putting it behind Decaf, FRONT, and Interspace in terms of performance, though it still outperforms BAP, WTF-PAD, and Spring. 

On the other hand, WF-DeTorrent performs only slightly worse than the FC setting, with a TPR of about .32 for a FPR of $10^{-4}$ and a TPR of about .58 for a $10^{-3}$ FPR, compared to FC-DeTorrent's TPRs of .30 and .57 respectively. Therefore, we find that WF-DeTorrent is transferable in the sense that it functions as an effective defense in both the WF and FC settings; accordingly, it may be a reasonable choice for users concerned about both types of attacks.

\subsection{Overhead}

While DeTorrent doesn't add any latency of its own, past work by Witwer et al. on network-wide Tor simulation \cite{paddingonly} has demonstrated that even zero-delay defenses likely strain the Tor network and incur latency. Their study used a `time-to-nth-byte' metric, finding that defenses Spring and Interspace had median latency overheads of 1.6\% and 7.8\% for 1MiB downloads compared to undefended downloads. Since the Spring and Interspace defenses avoid delaying traffic and incur similar bandwidth overhead compared to DeTorrent, as shown in Tables \ref{tab:df_overhead} and \ref{tab:be_overhead}, we might expect that DeTorrent would have similar results, though low observed bandwidth overhead and high failure rates in the experiments make precise estimation difficult. Still, it's quite possible that widespread adoption of DeTorrent would lead to non-negligible latency overhead.

\subsection{Implementation Framework}
While the PT system is convenient due to its relative flexibility in terms of supporting various traffic analysis defenses \cite{pt}, it only obfuscates the traffic between the client and the bridge. While this protects against local eavesdroppers, it leaves the user vulnerable to malicious bridges. 

Tor has implemented a circuit padding framework \cite{padding}, which adds dummy cells into circuit traffic to protect against WF attacks. Because the padding can be sent between the client and any relay in the circuit, it can be configured to stop at a middle node and thus protect against a malicious guard. However, the circuit padding framework is limited as it cannot delay traffic and must add dummy traffic based on pre-defined state machines. As a result, it is inflexible in terms of defense implementation and does not natively support randomized padding or complex pattern recognition, preventing it from being able to support DeTorrent.

\section{Conclusion and Future Work}

In this paper, we present the padding-only traffic analysis defense DeTorrent, which uses competing neural networks to generate defense strategies resistant to adversarial retraining. To train the defense generator to make real-time decisions, we implement it as an LSTM that takes into account live traffic patterns. We describe how we represent the trace as a tensor and use an embedder to create a useful notion of distance between traces, allowing for more effective training in the WF setting.

To test WF-DeTorrent's defense performance, we compare it to other padding-only defenses scaled to use similar bandwidth overhead. In the closed-world WF attack on the BE dataset, we find that it outperforms comparable defenses with a 10.5\% larger accuracy reduction than the next best defense. It also reduces the accuracy by 61.5\% compared to the undefended traffic, indicating its ability to defend against state-of-the-art WF attacks. Similarly, we find that FC-DeTorrent reduces the true positive rate of the flow correlation attack DeepCoFFEA at the $10^{-4}$ false positive rate threshold to just .30 compared to about .54 for the next best defense. Furthermore, we test the real-world DeTorrent implementation against live traffic and find that it achieves similar success, reducing Tik-Tok attack performance against the PT dataset to just 21.5\%. Overall, the results show that DeTorrent can both defend against traffic analysis attacks and be deployed practically. Future work may include experimenting with alternate neural network architectures or testing how delaying packets improves defense performance.

\section{Appendices}



\begin{acks}
We'd like to thank Nate Matthews for sharing the BigEnough dataset and for his work on tor-crawler-pluggable, which we used to collect the PT dataset. This work was funded by a 3M fellowship and NSF grants 1814753 and 1815757. It was also supported by the National Research Foundation of Korea (NRF) grant funded by the Korean government (MSIT) (No. RS-2022-00166669, RS-2023-00222385) and Institute of Information \& Communications Technology Planning \& Evaluation (IITP) grant funded by the Korean government (MSIT) (No. RS-2022-00155966, Artificial Intelligence Convergence Innovation Human Resources Development (Ewha Womans University)). 

This project was partially sponsored by the National Security Agency under Grant H98230-22-1-0302. The United States Government is authorized to reproduce and distribute reprints notwithstanding any copyright notation herein. Any opinions, findings, and conclusions or recommendations expressed in this material are those of the author(s) and do not necessarily reflect the views of the National Security Agency.
\end{acks}

\bibliographystyle{ACM-Reference-Format}
\bibliography{reg_bib}


\begin{thebibliography}{84}


\ifx \showCODEN    \undefined \def \showCODEN     #1{\unskip}     \fi
\ifx \showDOI      \undefined \def \showDOI       #1{#1}\fi
\ifx \showISBNx    \undefined \def \showISBNx     #1{\unskip}     \fi
\ifx \showISBNxiii \undefined \def \showISBNxiii  #1{\unskip}     \fi
\ifx \showISSN     \undefined \def \showISSN      #1{\unskip}     \fi
\ifx \showLCCN     \undefined \def \showLCCN      #1{\unskip}     \fi
\ifx \shownote     \undefined \def \shownote      #1{#1}          \fi
\ifx \showarticletitle \undefined \def \showarticletitle #1{#1}   \fi
\ifx \showURL      \undefined \def \showURL       {\relax}        \fi
\providecommand\bibfield[2]{#2}
\providecommand\bibinfo[2]{#2}
\providecommand\natexlab[1]{#1}
\providecommand\showeprint[2][]{arXiv:#2}

\bibitem[Abe and Goto(2016)]%
        {Abe2016}
\bibfield{author}{\bibinfo{person}{Kota Abe} {and} \bibinfo{person}{Shigeki
  Goto}.} \bibinfo{year}{2016}\natexlab{}.
\newblock \showarticletitle{{Fingerprinting Attack on Tor Anonymity using Deep
  Learning}}.
\newblock \bibinfo{journal}{\emph{Proceedings of the Asia-Pacific Advanced
  Network}} \bibinfo{volume}{42}, \bibinfo{number}{0} (\bibinfo{year}{2016}),
  \bibinfo{pages}{15--20}.
\newblock
\showISBNx{9784990544867}
\showISSN{2227-3026}


\bibitem[Akhoondi et~al\mbox{.}(2014)]%
        {akhoondi2012}
\bibfield{author}{\bibinfo{person}{Masoud Akhoondi}, \bibinfo{person}{Curtis
  Yu}, {and} \bibinfo{person}{Harsha~V. Madhyastha}.}
  \bibinfo{year}{2014}\natexlab{}.
\newblock \showarticletitle{LASTor: A Low-Latency AS-Aware Tor Client}.
\newblock \bibinfo{journal}{\emph{IEEE/ACM Trans. Netw.}} \bibinfo{volume}{22},
  \bibinfo{number}{6} (\bibinfo{date}{dec} \bibinfo{year}{2014}),
  \bibinfo{pages}{1742–1755}.
\newblock
\showISSN{1063-6692}
\urldef\tempurl%
\url{https://doi.org/10.1109/TNET.2013.2291242}
\showDOI{\tempurl}


\bibitem[Amazon(2022)]%
        {alexa}
\bibfield{author}{\bibinfo{person}{Amazon}.} \bibinfo{year}{2022}\natexlab{}.
\newblock \bibinfo{title}{Alexa top 1m}.
\newblock
  \bibinfo{howpublished}{\url{http://s3.amazonaws.com/alexa-static/top-1m.csv.zip}}.
\newblock
\newblock
\shownote{Accessed: 2022-08-01}.


\bibitem[Arjovsky et~al\mbox{.}(2017)]%
        {wasserstein}
\bibfield{author}{\bibinfo{person}{Martin Arjovsky}, \bibinfo{person}{Soumith
  Chintala}, {and} \bibinfo{person}{L{\'e}on Bottou}.}
  \bibinfo{year}{2017}\natexlab{}.
\newblock \showarticletitle{{W}asserstein Generative Adversarial Networks}. In
  \bibinfo{booktitle}{\emph{Proceedings of the 34th International Conference on
  Machine Learning}} \emph{(\bibinfo{series}{Proceedings of Machine Learning
  Research}, Vol.~\bibinfo{volume}{70})},
  \bibfield{editor}{\bibinfo{person}{Doina Precup} {and}
  \bibinfo{person}{Yee~Whye Teh}} (Eds.). \bibinfo{publisher}{PMLR},
  \bibinfo{pages}{214--223}.
\newblock
\urldef\tempurl%
\url{https://proceedings.mlr.press/v70/arjovsky17a.html}
\showURL{%
\tempurl}


\bibitem[arma(2011)]%
        {rotation}
\bibfield{author}{\bibinfo{person}{arma}.} \bibinfo{year}{2011}\natexlab{}.
\newblock \bibinfo{title}{Research problem: better guard rotation parameters}.
\newblock
\newblock
\urldef\tempurl%
\url{https://blog.torproject.org/research-problem-better-guard-rotation-parameters/}
\showURL{%
\tempurl}
\newblock
\shownote{Accessed: 2022-7-14}.


\bibitem[Barton and Wright(2016)]%
        {Armon2016}
\bibfield{author}{\bibinfo{person}{Armon Barton} {and} \bibinfo{person}{Matthew
  Wright}.} \bibinfo{year}{2016}\natexlab{}.
\newblock \showarticletitle{DeNASA: Destination-Naive AS-Awareness in Anonymous
  Communications}.
\newblock \bibinfo{journal}{\emph{Proceedings on Privacy Enhancing
  Technologies}}  \bibinfo{volume}{2016} (\bibinfo{date}{02}
  \bibinfo{year}{2016}).
\newblock
\urldef\tempurl%
\url{https://doi.org/10.1515/popets-2016-0044}
\showDOI{\tempurl}


\bibitem[Bhat et~al\mbox{.}(2019)]%
        {Bhat2019}
\bibfield{author}{\bibinfo{person}{Sanjit Bhat}, \bibinfo{person}{David Lu},
  \bibinfo{person}{Albert Kwon}, {and} \bibinfo{person}{Srinivas Devadas}.}
  \bibinfo{year}{2019}\natexlab{}.
\newblock \showarticletitle{{Var-CNN: A Data-Efficient Website Fingerprinting
  Attack Based on Deep Learning}}.
\newblock \bibinfo{journal}{\emph{Proceedings on Privacy Enhancing
  Technologies}} \bibinfo{volume}{2019}, \bibinfo{number}{4}
  (\bibinfo{date}{oct} \bibinfo{year}{2019}), \bibinfo{pages}{292--310}.
\newblock
\showISSN{2299-0984}
\urldef\tempurl%
\url{https://doi.org/10.2478/popets-2019-0070}
\showDOI{\tempurl}


\bibitem[Borisov et~al\mbox{.}(2007)]%
        {Borisov07}
\bibfield{author}{\bibinfo{person}{Nikita Borisov}, \bibinfo{person}{George
  Danezis}, \bibinfo{person}{Prateek Mittal}, {and} \bibinfo{person}{Parisa
  Tabriz}.} \bibinfo{year}{2007}\natexlab{}.
\newblock \showarticletitle{Denial of Service or Denial of Security? {H}ow
  Attacks on Reliability can Compromise Anonymity}. In
  \bibinfo{booktitle}{\emph{Proceedings of CCS 2007}}.
\newblock


\bibitem[Cai et~al\mbox{.}(2014a)]%
        {Cai2014a}
\bibfield{author}{\bibinfo{person}{Xiang Cai}, \bibinfo{person}{Rishab
  Nithyanand}, {and} \bibinfo{person}{Rob Johnson}.}
  \bibinfo{year}{2014}\natexlab{a}.
\newblock \showarticletitle{CS-BuFLO: A Congestion Sensitive Website
  Fingerprinting Defense}. In \bibinfo{booktitle}{\emph{Proceedings of the 13th
  Workshop on Privacy in the Electronic Society}} (Scottsdale, Arizona, USA)
  \emph{(\bibinfo{series}{WPES '14})}. \bibinfo{publisher}{Association for
  Computing Machinery}, \bibinfo{address}{New York, NY, USA},
  \bibinfo{pages}{121–130}.
\newblock
\showISBNx{9781450331487}
\urldef\tempurl%
\url{https://doi.org/10.1145/2665943.2665949}
\showDOI{\tempurl}


\bibitem[Cai et~al\mbox{.}(2014b)]%
        {Cai2014}
\bibfield{author}{\bibinfo{person}{Xiang Cai}, \bibinfo{person}{Rishab
  Nithyanand}, \bibinfo{person}{Tao Wang}, \bibinfo{person}{Rob Johnson}, {and}
  \bibinfo{person}{Ian Goldberg}.} \bibinfo{year}{2014}\natexlab{b}.
\newblock \showarticletitle{{A systematic approach to developing and evaluating
  website fingerprinting defenses}}. In \bibinfo{booktitle}{\emph{Proceedings
  of the ACM Conference on Computer and Communications Security}}.
  \bibinfo{publisher}{Association for Computing Machinery},
  \bibinfo{pages}{227--238}.
\newblock
\showISBNx{9781450329576}
\showISSN{15437221}
\urldef\tempurl%
\url{https://doi.org/10.1145/2660267.2660362}
\showDOI{\tempurl}


\bibitem[Cai et~al\mbox{.}(2012)]%
        {Cai2012}
\bibfield{author}{\bibinfo{person}{Xiang Cai}, \bibinfo{person}{Xin~Cheng
  Zhang}, \bibinfo{person}{Brijesh Joshi}, {and} \bibinfo{person}{Rob
  Johnson}.} \bibinfo{year}{2012}\natexlab{}.
\newblock \showarticletitle{Touching from a Distance: Website Fingerprinting
  Attacks and Defenses}. In \bibinfo{booktitle}{\emph{Proceedings of the 2012
  ACM Conference on Computer and Communications Security}} (Raleigh, North
  Carolina, USA) \emph{(\bibinfo{series}{CCS '12})}.
  \bibinfo{publisher}{Association for Computing Machinery},
  \bibinfo{address}{New York, NY, USA}, \bibinfo{pages}{605–616}.
\newblock
\showISBNx{9781450316514}
\urldef\tempurl%
\url{https://doi.org/10.1145/2382196.2382260}
\showDOI{\tempurl}


\bibitem[Crovella and Bestavros(1997)]%
        {selfsimilar}
\bibfield{author}{\bibinfo{person}{M.E. Crovella} {and} \bibinfo{person}{A.
  Bestavros}.} \bibinfo{year}{1997}\natexlab{}.
\newblock \showarticletitle{Self-similarity in World Wide Web traffic: evidence
  and possible causes}.
\newblock \bibinfo{journal}{\emph{IEEE/ACM Transactions on Networking}}
  \bibinfo{volume}{5}, \bibinfo{number}{6} (\bibinfo{year}{1997}),
  \bibinfo{pages}{835--846}.
\newblock
\urldef\tempurl%
\url{https://doi.org/10.1109/90.650143}
\showDOI{\tempurl}


\bibitem[De~la Cadena et~al\mbox{.}(2020)]%
        {trafficsliver}
\bibfield{author}{\bibinfo{person}{Wladimir De~la Cadena},
  \bibinfo{person}{Asya Mitseva}, \bibinfo{person}{Jens Hiller},
  \bibinfo{person}{Jan Pennekamp}, \bibinfo{person}{Sebastian Reuter},
  \bibinfo{person}{Julian Filter}, \bibinfo{person}{Thomas Engel},
  \bibinfo{person}{Klaus Wehrle}, {and} \bibinfo{person}{Andriy Panchenko}.}
  \bibinfo{year}{2020}\natexlab{}.
\newblock \showarticletitle{TrafficSliver: Fighting Website Fingerprinting
  Attacks with Traffic Splitting}. In \bibinfo{booktitle}{\emph{Proceedings of
  the 2020 ACM SIGSAC Conference on Computer and Communications Security}}
  (Virtual Event, USA) \emph{(\bibinfo{series}{CCS '20})}.
  \bibinfo{publisher}{Association for Computing Machinery},
  \bibinfo{address}{New York, NY, USA}, \bibinfo{pages}{1971–1985}.
\newblock
\showISBNx{9781450370899}
\urldef\tempurl%
\url{https://doi.org/10.1145/3372297.3423351}
\showDOI{\tempurl}


\bibitem[Dingledine et~al\mbox{.}(2004)]%
        {Dingledine2004}
\bibfield{author}{\bibinfo{person}{Roger Dingledine}, \bibinfo{person}{Nick
  Mathewson}, {and} \bibinfo{person}{Paul Syverson}.}
  \bibinfo{year}{2004}\natexlab{}.
\newblock \showarticletitle{Tor: The {Second-Generation} Onion Router}. In
  \bibinfo{booktitle}{\emph{13th USENIX Security Symposium (USENIX Security
  04)}}. \bibinfo{publisher}{USENIX Association}, \bibinfo{address}{San Diego,
  CA}.
\newblock
\urldef\tempurl%
\url{https://www.usenix.org/conference/13th-usenix-security-symposium/tor-second-generation-onion-router}
\showURL{%
\tempurl}


\bibitem[Dyer et~al\mbox{.}(2012)]%
        {Dyer2012}
\bibfield{author}{\bibinfo{person}{Kevin~P. Dyer}, \bibinfo{person}{Scott~E.
  Coull}, \bibinfo{person}{Thomas Ristenpart}, {and} \bibinfo{person}{Thomas
  Shrimpton}.} \bibinfo{year}{2012}\natexlab{}.
\newblock \showarticletitle{Peek-a-Boo, I Still See You: Why Efficient Traffic
  Analysis Countermeasures Fail}. In \bibinfo{booktitle}{\emph{2012 IEEE
  Symposium on Security and Privacy}}. \bibinfo{pages}{332--346}.
\newblock
\urldef\tempurl%
\url{https://doi.org/10.1109/SP.2012.28}
\showDOI{\tempurl}


\bibitem[Edman and Syverson(2009)]%
        {edman2009}
\bibfield{author}{\bibinfo{person}{Matthew Edman} {and} \bibinfo{person}{Paul
  Syverson}.} \bibinfo{year}{2009}\natexlab{}.
\newblock \showarticletitle{As-Awareness in Tor Path Selection}. In
  \bibinfo{booktitle}{\emph{Proceedings of the 16th ACM Conference on Computer
  and Communications Security}} (Chicago, Illinois, USA)
  \emph{(\bibinfo{series}{CCS '09})}. \bibinfo{publisher}{Association for
  Computing Machinery}, \bibinfo{address}{New York, NY, USA},
  \bibinfo{pages}{380–389}.
\newblock
\showISBNx{9781605588940}
\urldef\tempurl%
\url{https://doi.org/10.1145/1653662.1653708}
\showDOI{\tempurl}


\bibitem[Feamster and Dingledine(2004)]%
        {feamster2004}
\bibfield{author}{\bibinfo{person}{Nick Feamster} {and} \bibinfo{person}{Roger
  Dingledine}.} \bibinfo{year}{2004}\natexlab{}.
\newblock \showarticletitle{Location Diversity in Anonymity Networks}. In
  \bibinfo{booktitle}{\emph{Proceedings of the 2004 ACM Workshop on Privacy in
  the Electronic Society}} (Washington DC, USA) \emph{(\bibinfo{series}{WPES
  '04})}. \bibinfo{publisher}{Association for Computing Machinery},
  \bibinfo{address}{New York, NY, USA}, \bibinfo{pages}{66–76}.
\newblock
\showISBNx{1581139683}
\urldef\tempurl%
\url{https://doi.org/10.1145/1029179.1029199}
\showDOI{\tempurl}


\bibitem[Gong and Wang(2020)]%
        {Gong2020}
\bibfield{author}{\bibinfo{person}{Jiajun Gong} {and} \bibinfo{person}{Tao
  Wang}.} \bibinfo{year}{2020}\natexlab{}.
\newblock \showarticletitle{Zero-delay Lightweight Defenses against Website
  Fingerprinting}. In \bibinfo{booktitle}{\emph{29th USENIX Security Symposium
  (USENIX Security 20)}}. \bibinfo{publisher}{USENIX Association},
  \bibinfo{pages}{717--734}.
\newblock
\showISBNx{978-1-939133-17-5}
\urldef\tempurl%
\url{https://www.usenix.org/conference/usenixsecurity20/presentation/gong}
\showURL{%
\tempurl}


\bibitem[Gong and Wang(2022)]%
        {frontcode}
\bibfield{author}{\bibinfo{person}{Jiajun Gong} {and} \bibinfo{person}{Tao
  Wang}.} \bibinfo{year}{2022}\natexlab{}.
\newblock \bibinfo{title}{\textit{WebsiteFingerprinting}}.
\newblock
  \bibinfo{howpublished}{\url{https://github.com/websitefingerprinting/WebsiteFingerprinting}}.
\newblock
\newblock
\shownote{Accessed: 2022-7-14}.


\bibitem[Gong et~al\mbox{.}(2022)]%
        {surakav}
\bibfield{author}{\bibinfo{person}{Jiajun Gong}, \bibinfo{person}{Wuqi Zhang},
  \bibinfo{person}{Charles Zhang}, {and} \bibinfo{person}{Tao Wang}.}
  \bibinfo{year}{2022}\natexlab{}.
\newblock \showarticletitle{Surakav: Generating Realistic Traces for a Strong
  Website Fingerprinting Defense}. In \bibinfo{booktitle}{\emph{2022 IEEE
  Symposium on Security and Privacy (SP)}}. \bibinfo{publisher}{IEEE},
  \bibinfo{pages}{1558--1573}.
\newblock
\urldef\tempurl%
\url{https://doi.org/10.1109/SP46214.2022.9833722}
\showDOI{\tempurl}


\bibitem[Goodfellow et~al\mbox{.}(2014)]%
        {gan}
\bibfield{author}{\bibinfo{person}{Ian Goodfellow}, \bibinfo{person}{Jean
  Pouget-Abadie}, \bibinfo{person}{Mehdi Mirza}, \bibinfo{person}{Bing Xu},
  \bibinfo{person}{David Warde-Farley}, \bibinfo{person}{Sherjil Ozair},
  \bibinfo{person}{Aaron Courville}, {and} \bibinfo{person}{Yoshua Bengio}.}
  \bibinfo{year}{2014}\natexlab{}.
\newblock \showarticletitle{Generative Adversarial Nets}. In
  \bibinfo{booktitle}{\emph{Advances in Neural Information Processing
  Systems}}, \bibfield{editor}{\bibinfo{person}{Z.~Ghahramani},
  \bibinfo{person}{M.~Welling}, \bibinfo{person}{C.~Cortes},
  \bibinfo{person}{N.~Lawrence}, {and} \bibinfo{person}{K.Q. Weinberger}}
  (Eds.), Vol.~\bibinfo{volume}{27}. \bibinfo{publisher}{Curran Associates,
  Inc.}
\newblock
\urldef\tempurl%
\url{https://proceedings.neurips.cc/paper_files/paper/2014/file/5ca3e9b122f61f8f06494c97b1afccf3-Paper.pdf}
\showURL{%
\tempurl}


\bibitem[Hayes and Danezis(2016)]%
        {Hayes2015}
\bibfield{author}{\bibinfo{person}{Jamie Hayes} {and} \bibinfo{person}{George
  Danezis}.} \bibinfo{year}{2016}\natexlab{}.
\newblock \showarticletitle{k-fingerprinting: A Robust Scalable Website
  Fingerprinting Technique}. In \bibinfo{booktitle}{\emph{25th USENIX Security
  Symposium (USENIX Security 16)}}. \bibinfo{publisher}{USENIX Association},
  \bibinfo{address}{Austin, TX}, \bibinfo{pages}{1187--1203}.
\newblock
\showISBNx{978-1-931971-32-4}
\urldef\tempurl%
\url{https://www.usenix.org/conference/usenixsecurity16/technical-sessions/presentation/hayes}
\showURL{%
\tempurl}


\bibitem[Henri et~al\mbox{.}(2020)]%
        {multihoming}
\bibfield{author}{\bibinfo{person}{S{\'e}bastien Henri}, \bibinfo{person}{Gines
  Garcia-Aviles}, \bibinfo{person}{P. Serrano}, \bibinfo{person}{A. Banchs},
  {and} \bibinfo{person}{P. Thiran}.} \bibinfo{year}{2020}\natexlab{}.
\newblock \showarticletitle{Protecting against Website Fingerprinting with
  Multihoming}.
\newblock \bibinfo{journal}{\emph{Proceedings on Privacy Enhancing
  Technologies}}  \bibinfo{volume}{2020} (\bibinfo{year}{2020}),
  \bibinfo{pages}{89 -- 110}.
\newblock


\bibitem[Herrmann et~al\mbox{.}(2009)]%
        {Herrmann2009}
\bibfield{author}{\bibinfo{person}{Dominik Herrmann}, \bibinfo{person}{Rolf
  Wendolsky}, {and} \bibinfo{person}{Hannes Federrath}.}
  \bibinfo{year}{2009}\natexlab{}.
\newblock \showarticletitle{Website Fingerprinting: Attacking Popular Privacy
  Enhancing Technologies with the Multinomial Na\"{\i}ve-Bayes Classifier}. In
  \bibinfo{booktitle}{\emph{Proceedings of the 2009 ACM Workshop on Cloud
  Computing Security}} (Chicago, Illinois, USA) \emph{(\bibinfo{series}{CCSW
  '09})}. \bibinfo{publisher}{Association for Computing Machinery},
  \bibinfo{address}{New York, NY, USA}, \bibinfo{pages}{31–42}.
\newblock
\showISBNx{9781605587844}
\urldef\tempurl%
\url{https://doi.org/10.1145/1655008.1655013}
\showDOI{\tempurl}


\bibitem[Hintz(2002)]%
        {Hintz2002}
\bibfield{author}{\bibinfo{person}{Andrew Hintz}.}
  \bibinfo{year}{2002}\natexlab{}.
\newblock \showarticletitle{Fingerprinting Websites Using Traffic Analysis}. In
  \bibinfo{booktitle}{\emph{Proceedings of the 2nd International Conference on
  Privacy Enhancing Technologies}} (San Francisco, CA, USA)
  \emph{(\bibinfo{series}{PET'02})}. \bibinfo{publisher}{Springer-Verlag},
  \bibinfo{address}{Berlin, Heidelberg}, \bibinfo{pages}{171–178}.
\newblock
\showISBNx{354000565X}


\bibitem[Hochreiter and Schmidhuber(1997)]%
        {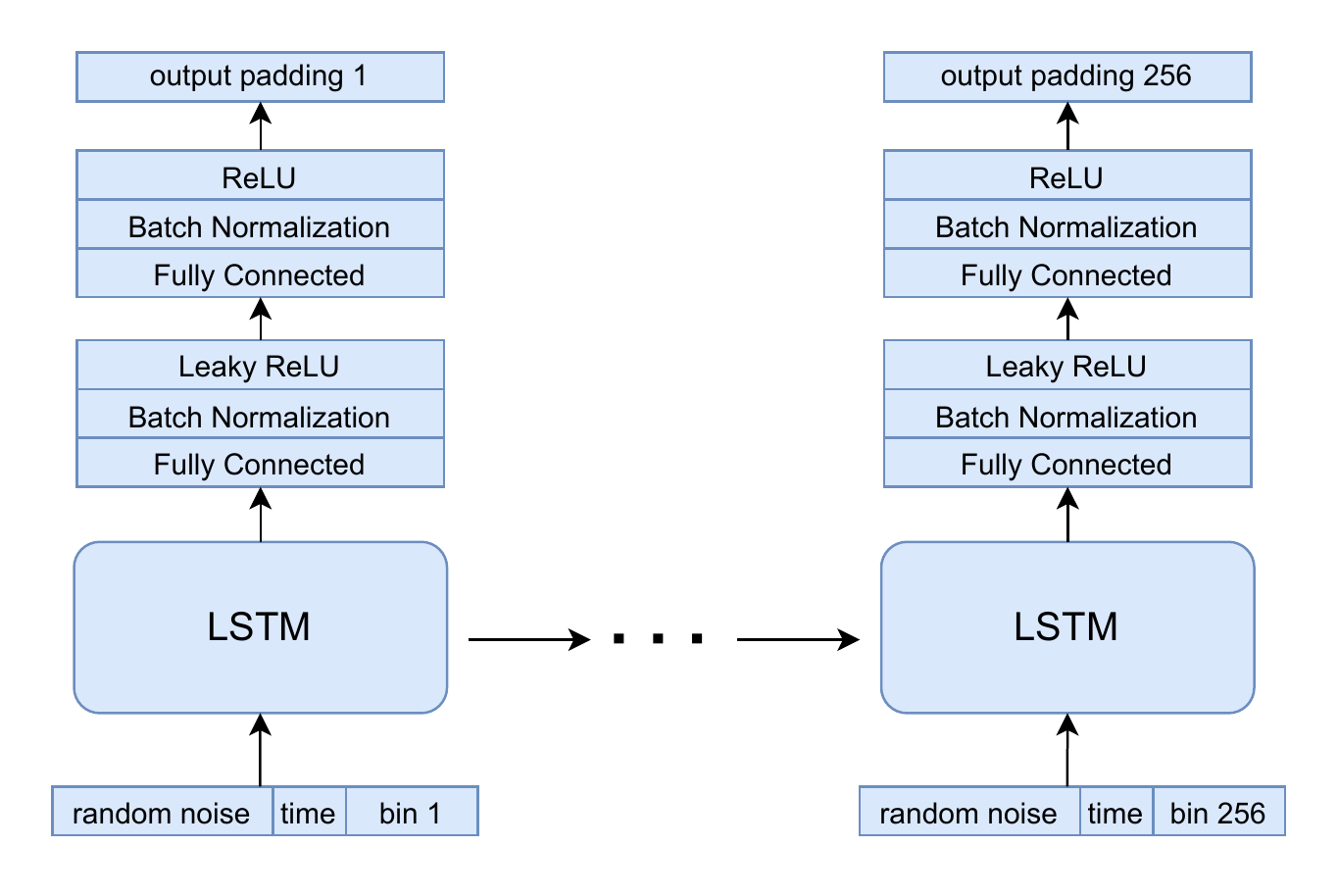}
\bibfield{author}{\bibinfo{person}{Sepp Hochreiter} {and}
  \bibinfo{person}{Jürgen Schmidhuber}.} \bibinfo{year}{1997}\natexlab{}.
\newblock \showarticletitle{{Long Short-Term Memory}}.
\newblock \bibinfo{journal}{\emph{Neural Computation}} \bibinfo{volume}{9},
  \bibinfo{number}{8} (\bibinfo{date}{11} \bibinfo{year}{1997}),
  \bibinfo{pages}{1735--1780}.
\newblock
\showISSN{0899-7667}
\urldef\tempurl%
\url{https://doi.org/10.1162/neco.1997.9.8.1735}
\showDOI{\tempurl}
\showeprint{https://direct.mit.edu/neco/article-pdf/9/8/1735/813796/neco.1997.9.8.1735.pdf}


\bibitem[Holland and Hopper(2022)]%
        {holland2022}
\bibfield{author}{\bibinfo{person}{James~K Holland} {and}
  \bibinfo{person}{Nicholas Hopper}.} \bibinfo{year}{2022}\natexlab{}.
\newblock \showarticletitle{{RegulaTor: A Straightforward Website
  Fingerprinting Defense}}.
\newblock \bibinfo{journal}{\emph{Proceedings on Privacy Enhancing
  Technologies}}  \bibinfo{volume}{2022} (\bibinfo{year}{2022}).
\newblock
\urldef\tempurl%
\url{https://petsymposium.org/popets/2022/popets-2022-0049.pdf}
\showURL{%
\tempurl}


\bibitem[Houmansadr et~al\mbox{.}(2012)]%
        {watermarking}
\bibfield{author}{\bibinfo{person}{Amir Houmansadr}, \bibinfo{person}{Negar
  Kiyavash}, {and} \bibinfo{person}{Nikita Borisov}.}
  \bibinfo{year}{2012}\natexlab{}.
\newblock \showarticletitle{Non-Blind Watermarking of Network Flows}.
\newblock \bibinfo{journal}{\emph{IEEE/ACM Transactions on Networking}}
  \bibinfo{volume}{22} (\bibinfo{date}{03} \bibinfo{year}{2012}).
\newblock
\urldef\tempurl%
\url{https://doi.org/10.1109/TNET.2013.2272740}
\showDOI{\tempurl}


\bibitem[Initiative(2022)]%
        {pagerank}
\bibfield{author}{\bibinfo{person}{Open~PageRank Initiative}.}
  \bibinfo{year}{2022}\natexlab{}.
\newblock \bibinfo{title}{What is Open PageRank?}
\newblock
  \bibinfo{howpublished}{\url{https://www.domcop.com/openpagerank/what-is-openpagerank}}.
\newblock
\newblock
\shownote{Accessed: 2022-08-01}.


\bibitem[Jiang and Dovrolis(2005)]%
        {burstyshort}
\bibfield{author}{\bibinfo{person}{Hao Jiang} {and}
  \bibinfo{person}{Constantinos Dovrolis}.} \bibinfo{year}{2005}\natexlab{}.
\newblock \showarticletitle{Why is the Internet Traffic Bursty in Short Time
  Scales?}
\newblock \bibinfo{journal}{\emph{SIGMETRICS Perform. Eval. Rev.}}
  \bibinfo{volume}{33}, \bibinfo{number}{1} (\bibinfo{date}{jun}
  \bibinfo{year}{2005}), \bibinfo{pages}{241–252}.
\newblock
\showISSN{0163-5999}
\urldef\tempurl%
\url{https://doi.org/10.1145/1071690.1064240}
\showDOI{\tempurl}


\bibitem[Johnson et~al\mbox{.}(2013)]%
        {johnson2013}
\bibfield{author}{\bibinfo{person}{Aaron Johnson}, \bibinfo{person}{Chris
  Wacek}, \bibinfo{person}{Rob Jansen}, \bibinfo{person}{Micah Sherr}, {and}
  \bibinfo{person}{Paul Syverson}.} \bibinfo{year}{2013}\natexlab{}.
\newblock \showarticletitle{Users Get Routed: Traffic Correlation on Tor by
  Realistic Adversaries}. In \bibinfo{booktitle}{\emph{Proceedings of the 2013
  ACM SIGSAC Conference on Computer and Communications Security}} (Berlin,
  Germany) \emph{(\bibinfo{series}{CCS '13})}. \bibinfo{publisher}{Association
  for Computing Machinery}, \bibinfo{address}{New York, NY, USA},
  \bibinfo{pages}{337–348}.
\newblock
\showISBNx{9781450324779}
\urldef\tempurl%
\url{https://doi.org/10.1145/2508859.2516651}
\showDOI{\tempurl}


\bibitem[Juarez(2015)]%
        {wfpadtools}
\bibfield{author}{\bibinfo{person}{Marc Juarez}.}
  \bibinfo{year}{2015}\natexlab{}.
\newblock \bibinfo{title}{WFPadTools}.
\newblock \bibinfo{howpublished}{\url{https://github.com/mjuarezm/wfpadtools}}.
\newblock
\newblock
\shownote{Accessed: 2021-5-21}.


\bibitem[Juarez et~al\mbox{.}(2015)]%
        {Juarez2015}
\bibfield{author}{\bibinfo{person}{Marc Juarez}, \bibinfo{person}{Mohsen
  Imani}, \bibinfo{person}{Mike Perry}, \bibinfo{person}{Claudia Diaz}, {and}
  \bibinfo{person}{Matthew Wright}.} \bibinfo{year}{2015}\natexlab{}.
\newblock \showarticletitle{{WTF-PAD: Toward an Efficient Website
  Fingerprinting Defense for Tor}}.
\newblock \bibinfo{journal}{\emph{Computing Research Repository (CoRR)}}
  \bibinfo{volume}{abs/1512.0} (\bibinfo{year}{2015}).
\newblock
\showISBNx{978-3-319-45744-4}
\showeprint[arxiv]{1512.00524}
\urldef\tempurl%
\url{http://arxiv.org/abs/1512.00524}
\showURL{%
\tempurl}


\bibitem[Juarez et~al\mbox{.}(2022)]%
        {wtfpadcode}
\bibfield{author}{\bibinfo{person}{M. Juarez}, \bibinfo{person}{M. Imani},
  \bibinfo{person}{M. Perry}, \bibinfo{person}{C. Diaz}, {and}
  \bibinfo{person}{M. Wright}.} \bibinfo{year}{2022}\natexlab{}.
\newblock \bibinfo{title}{WTF-PAD}.
\newblock
\newblock
\urldef\tempurl%
\url{https://github.com/wtfpad/wtfpad}
\showURL{%
\tempurl}
\newblock
\shownote{Accessed: 2022-7-14}.


\bibitem[Juen et~al\mbox{.}(2014)]%
        {juen2014}
\bibfield{author}{\bibinfo{person}{Joshua Juen}, \bibinfo{person}{Anupam Das},
  \bibinfo{person}{Aaron Johnson}, \bibinfo{person}{Nikita Borisov}, {and}
  \bibinfo{person}{Matthew Caesar}.} \bibinfo{year}{2014}\natexlab{}.
\newblock \showarticletitle{Defending Tor from Network Adversaries: {A} Case
  Study of Network Path Prediction}.
\newblock \bibinfo{journal}{\emph{CoRR}}  \bibinfo{volume}{abs/1410.1823}
  (\bibinfo{year}{2014}).
\newblock
\showeprint[arXiv]{1410.1823}
\urldef\tempurl%
\url{http://arxiv.org/abs/1410.1823}
\showURL{%
\tempurl}


\bibitem[Karonen(2014)]%
        {packetrate}
\bibfield{author}{\bibinfo{person}{Ilmari Karonen}.}
  \bibinfo{year}{2014}\natexlab{}.
\newblock \bibinfo{title}{Estimating rate of occurrence of an event with
  exponential smoothing and irregular events}.
\newblock
\newblock
\urldef\tempurl%
\url{https://stackoverflow.com/questions/23615974/estimating-rate-of-occurrence-of-an-event-with-exponential-smoothing-and-irregul?noredirect=1&lq=1}
\showURL{%
\tempurl}
\newblock
\shownote{Accessed: 2022-8-13}.


\bibitem[Levine et~al\mbox{.}(2004)]%
        {Levine2004}
\bibfield{author}{\bibinfo{person}{Brian~N. Levine},
  \bibinfo{person}{Michael~K. Reiter}, \bibinfo{person}{Chenxi Wang}, {and}
  \bibinfo{person}{Matthew Wright}.} \bibinfo{year}{2004}\natexlab{}.
\newblock \showarticletitle{Timing Attacks in Low-Latency Mix Systems}. In
  \bibinfo{booktitle}{\emph{Financial Cryptography}},
  \bibfield{editor}{\bibinfo{person}{Ari Juels}} (Ed.).
  \bibinfo{publisher}{Springer Berlin Heidelberg}, \bibinfo{address}{Berlin,
  Heidelberg}, \bibinfo{pages}{251--265}.
\newblock
\showISBNx{978-3-540-27809-2}


\bibitem[Luo et~al\mbox{.}(2011)]%
        {Luo2011}
\bibfield{author}{\bibinfo{person}{Xiapu Luo}, \bibinfo{person}{Peng Zhou},
  \bibinfo{person}{Edmond W~W Chan}, \bibinfo{person}{Wenke Lee},
  \bibinfo{person}{Rocky K~C Chang}, {and} \bibinfo{person}{Roberto Perdisci}.}
  \bibinfo{year}{2011}\natexlab{}.
\newblock \bibinfo{booktitle}{\emph{{HTTPOS: Sealing Information Leaks with
  Browser-side Obfuscation of Encrypted Flows}}}.
\newblock \bibinfo{type}{{T}echnical {R}eport}.
\newblock


\bibitem[Mani et~al\mbox{.}(2018)]%
        {measurement}
\bibfield{author}{\bibinfo{person}{Akshaya Mani}, \bibinfo{person}{T.
  Wilson-Brown}, \bibinfo{person}{Rob Jansen}, \bibinfo{person}{Aaron Johnson},
  {and} \bibinfo{person}{Micah Sherr}.} \bibinfo{year}{2018}\natexlab{}.
\newblock \showarticletitle{Understanding Tor Usage with Privacy-Preserving
  Measurement}. In \bibinfo{booktitle}{\emph{Proceedings of the Internet
  Measurement Conference 2018}} (Boston, MA, USA) \emph{(\bibinfo{series}{IMC
  '18})}. \bibinfo{publisher}{Association for Computing Machinery},
  \bibinfo{address}{New York, NY, USA}, \bibinfo{pages}{175–187}.
\newblock
\showISBNx{9781450356190}
\urldef\tempurl%
\url{https://doi.org/10.1145/3278532.3278549}
\showDOI{\tempurl}


\bibitem[Matthews et~al\mbox{.}(2023)]%
        {evaluation}
\bibfield{author}{\bibinfo{person}{Nate Matthews}, \bibinfo{person}{James
  Holland}, \bibinfo{person}{Se~Eun Oh}, \bibinfo{person}{Mohammad~Saidur
  Rahman}, \bibinfo{person}{Matthew Wright}, {and} \bibinfo{person}{Nicholas
  Hopper}.} \bibinfo{year}{2023}\natexlab{}.
\newblock \showarticletitle{SoK: A Critical Evaluation of Efficient Website
  Fingerprinting Defenses}. In \bibinfo{booktitle}{\emph{2023 IEEE Symposium on
  Security and Privacy}}. IEEE.
\newblock


\bibitem[Murdoch and Danezis(2005)]%
        {murdoch2005}
\bibfield{author}{\bibinfo{person}{S.J. Murdoch} {and} \bibinfo{person}{G.
  Danezis}.} \bibinfo{year}{2005}\natexlab{}.
\newblock \showarticletitle{Low-cost traffic analysis of Tor}. In
  \bibinfo{booktitle}{\emph{2005 IEEE Symposium on Security and Privacy
  (S\&P'05)}}. \bibinfo{pages}{183--195}.
\newblock
\urldef\tempurl%
\url{https://doi.org/10.1109/SP.2005.12}
\showDOI{\tempurl}


\bibitem[Nasr et~al\mbox{.}(2018)]%
        {deepcorr}
\bibfield{author}{\bibinfo{person}{Milad Nasr}, \bibinfo{person}{Alireza
  Bahramali}, {and} \bibinfo{person}{Amir Houmansadr}.}
  \bibinfo{year}{2018}\natexlab{}.
\newblock \showarticletitle{{DeepCorr}}. In
  \bibinfo{booktitle}{\emph{Proceedings of the 2018 {ACM} {SIGSAC} Conference
  on Computer and Communications Security}}. \bibinfo{publisher}{{ACM}}.
\newblock
\urldef\tempurl%
\url{https://doi.org/10.1145/3243734.3243824}
\showDOI{\tempurl}


\bibitem[Nasr et~al\mbox{.}(2021)]%
        {banp}
\bibfield{author}{\bibinfo{person}{Milad Nasr}, \bibinfo{person}{Alireza
  Bahramali}, {and} \bibinfo{person}{Amir Houmansadr}.}
  \bibinfo{year}{2021}\natexlab{}.
\newblock \showarticletitle{Defeating {DNN-Based} Traffic Analysis Systems in
  {Real-Time} With Blind Adversarial Perturbations}. In
  \bibinfo{booktitle}{\emph{30th USENIX Security Symposium (USENIX Security
  21)}}. \bibinfo{publisher}{USENIX Association}, \bibinfo{pages}{2705--2722}.
\newblock
\showISBNx{978-1-939133-24-3}
\urldef\tempurl%
\url{https://www.usenix.org/conference/usenixsecurity21/presentation/nasr}
\showURL{%
\tempurl}


\bibitem[Nasr et~al\mbox{.}(2017)]%
        {compressive}
\bibfield{author}{\bibinfo{person}{Milad Nasr}, \bibinfo{person}{Amir
  Houmansadr}, {and} \bibinfo{person}{Arya Mazumdar}.}
  \bibinfo{year}{2017}\natexlab{}.
\newblock \showarticletitle{Compressive Traffic Analysis: A New Paradigm for
  Scalable Traffic Analysis}. In \bibinfo{booktitle}{\emph{Proceedings of the
  2017 ACM SIGSAC Conference on Computer and Communications Security}} (Dallas,
  Texas, USA) \emph{(\bibinfo{series}{CCS '17})}.
  \bibinfo{publisher}{Association for Computing Machinery},
  \bibinfo{address}{New York, NY, USA}, \bibinfo{pages}{2053–2069}.
\newblock
\showISBNx{9781450349468}
\urldef\tempurl%
\url{https://doi.org/10.1145/3133956.3134074}
\showDOI{\tempurl}


\bibitem[Nithyanand et~al\mbox{.}(2014)]%
        {Nithyanand2014}
\bibfield{author}{\bibinfo{person}{Rishab Nithyanand}, \bibinfo{person}{Xiang
  Cai}, {and} \bibinfo{person}{Rob Johnson}.} \bibinfo{year}{2014}\natexlab{}.
\newblock \showarticletitle{{Glove: A bespoke website fingerprinting defense}}.
  In \bibinfo{booktitle}{\emph{Proceedings of the ACM Conference on Computer
  and Communications Security}}. \bibinfo{publisher}{Association for Computing
  Machinery}, \bibinfo{pages}{131--134}.
\newblock
\showISBNx{9781450331487}
\showISSN{15437221}
\urldef\tempurl%
\url{https://doi.org/10.1145/2665943.2665950}
\showDOI{\tempurl}


\bibitem[Oh et~al\mbox{.}(2020)]%
        {gandalf}
\bibfield{author}{\bibinfo{person}{Se Oh}, \bibinfo{person}{Nate Mathews},
  \bibinfo{person}{Mohammad~Saidur Rahman}, \bibinfo{person}{Matthew Wright},
  {and} \bibinfo{person}{Nicholas Hopper}.} \bibinfo{year}{2020}\natexlab{}.
\newblock \showarticletitle{GANDaLF: GAN for Data-Limited Fingerprinting}.
\newblock \bibinfo{journal}{\emph{Proceedings on Privacy Enhancing
  Technologies}}  \bibinfo{volume}{2021}.
\newblock
\urldef\tempurl%
\url{https://doi.org/10.2478/popets-2021-0029}
\showDOI{\tempurl}


\bibitem[Oh et~al\mbox{.}(2019)]%
        {Oh2019}
\bibfield{author}{\bibinfo{person}{Se~Eun Oh}, \bibinfo{person}{Saikrishna
  Sunkam}, {and} \bibinfo{person}{Nicholas Hopper}.}
  \bibinfo{year}{2019}\natexlab{}.
\newblock \showarticletitle{{p1-FP: Extraction, Classification, and Prediction
  of Website Fingerprints with Deep Learning}}.
\newblock \bibinfo{journal}{\emph{Proceedings on Privacy Enhancing
  Technologies}} \bibinfo{volume}{2019}, \bibinfo{number}{3}
  (\bibinfo{year}{2019}), \bibinfo{pages}{191--209}.
\newblock
\urldef\tempurl%
\url{https://doi.org/10.2478/popets-2019-0043}
\showDOI{\tempurl}
\showeprint[arxiv]{arXiv:1711.03656v2}


\bibitem[Oh et~al\mbox{.}(2022)]%
        {dcf}
\bibfield{author}{\bibinfo{person}{Se~Eun Oh}, \bibinfo{person}{Taiji Yang},
  \bibinfo{person}{Nate Mathews}, \bibinfo{person}{James~K Holland},
  \bibinfo{person}{Mohammad~Saidur Rahman}, \bibinfo{person}{Nicholas Hopper},
  {and} \bibinfo{person}{Matthew Wright}.} \bibinfo{year}{2022}\natexlab{}.
\newblock \showarticletitle{DeepCoFFEA: Improved flow correlation attacks on
  Tor via metric learning and amplification}. In \bibinfo{booktitle}{\emph{2022
  IEEE Symposium on Security and Privacy (SP)}}. IEEE,
  \bibinfo{pages}{1915--1932}.
\newblock


\bibitem[Overlier and Syverson(2006)]%
        {overlier2006}
\bibfield{author}{\bibinfo{person}{L. Overlier} {and} \bibinfo{person}{P.
  Syverson}.} \bibinfo{year}{2006}\natexlab{}.
\newblock \showarticletitle{Locating hidden servers}. In
  \bibinfo{booktitle}{\emph{2006 IEEE Symposium on Security and Privacy
  (S\&P'06)}}. \bibinfo{pages}{15 pp.--114}.
\newblock
\urldef\tempurl%
\url{https://doi.org/10.1109/SP.2006.24}
\showDOI{\tempurl}


\bibitem[Panchenko et~al\mbox{.}(2017)]%
        {Panchenko2017}
\bibfield{author}{\bibinfo{person}{Andriy Panchenko}, \bibinfo{person}{Fabian
  Lanze}, \bibinfo{person}{Andreas Zinnen}, \bibinfo{person}{Martin Henze},
  \bibinfo{person}{Jan Pennekamp}, \bibinfo{person}{Klaus Wehrle}, {and}
  \bibinfo{person}{Thomas Engel}.} \bibinfo{year}{2017}\natexlab{}.
\newblock \showarticletitle{{Website Fingerprinting at Internet Scale}}.
\newblock  \bibinfo{number}{February} (\bibinfo{year}{2017}),
  \bibinfo{pages}{21--24}.
\newblock
\showISBNx{189156241X}
\urldef\tempurl%
\url{https://doi.org/10.14722/ndss.2016.23477}
\showDOI{\tempurl}


\bibitem[Panchenko et~al\mbox{.}(2011)]%
        {Panchenko2011}
\bibfield{author}{\bibinfo{person}{Andriy Panchenko}, \bibinfo{person}{Lukas
  Niessen}, \bibinfo{person}{Andreas Zinnen}, {and} \bibinfo{person}{Thomas
  Engel}.} \bibinfo{year}{2011}\natexlab{}.
\newblock \showarticletitle{Website Fingerprinting in Onion Routing Based
  Anonymization Networks}. In \bibinfo{booktitle}{\emph{Proceedings of the 10th
  Annual ACM Workshop on Privacy in the Electronic Society}} (Chicago,
  Illinois, USA) \emph{(\bibinfo{series}{WPES '11})}.
  \bibinfo{publisher}{Association for Computing Machinery},
  \bibinfo{address}{New York, NY, USA}, \bibinfo{pages}{103–114}.
\newblock
\showISBNx{9781450310024}
\urldef\tempurl%
\url{https://doi.org/10.1145/2046556.2046570}
\showDOI{\tempurl}


\bibitem[Perry and Kadianakis(2021)]%
        {padding}
\bibfield{author}{\bibinfo{person}{Mike Perry} {and} \bibinfo{person}{George
  Kadianakis}.} \bibinfo{year}{2021}\natexlab{}.
\newblock \bibinfo{title}{Circuit Padding Developer Documentation}.
\newblock
  \bibinfo{howpublished}{\url{https://github.com/torproject/tor/blob/master/doc/HACKING/CircuitPaddingDevelopment.md}}.
\newblock
\newblock
\shownote{Accessed: 2022-09-17}.


\bibitem[Project(2014)]%
        {obfsproxy}
\bibfield{author}{\bibinfo{person}{Tor Project}.}
  \bibinfo{year}{2014}\natexlab{}.
\newblock \bibinfo{title}{Pluggable transport for obfuscated traffic}.
\newblock
  \bibinfo{howpublished}{\url{https://gitweb.torproject.org/pluggable-transports/obfsproxy.git}}.
\newblock
\newblock
\shownote{Accessed: 2022-7-21}.


\bibitem[Project(2022a)]%
        {bridge}
\bibfield{author}{\bibinfo{person}{Tor Project}.}
  \bibinfo{year}{2022}\natexlab{a}.
\newblock \bibinfo{title}{Get Bridges for Tor}.
\newblock
\newblock
\urldef\tempurl%
\url{https://bridges.torproject.org/}
\showURL{%
\tempurl}
\newblock
\shownote{Accessed: 2022-7-28}.


\bibitem[Project(2022b)]%
        {metrics}
\bibfield{author}{\bibinfo{person}{Tor Project}.}
  \bibinfo{year}{2022}\natexlab{b}.
\newblock \bibinfo{title}{Tor Metrics}.
\newblock
  \bibinfo{howpublished}{\url{https://metrics.torproject.org/userstats-relay-country.html}}.
\newblock
\newblock
\shownote{Accessed: 2022-09-20}.


\bibitem[Project(2022c)]%
        {pt}
\bibfield{author}{\bibinfo{person}{Tor Project}.}
  \bibinfo{year}{2022}\natexlab{c}.
\newblock \bibinfo{title}{Tor: Pluggable Transports}.
\newblock
  \bibinfo{howpublished}{\url{https://2019.www.torproject.org/docs/pluggable-transports.html.en}}.
\newblock
\newblock
\shownote{Accessed: 2022-5-21}.


\bibitem[Pulls(2019)]%
        {pullscode}
\bibfield{author}{\bibinfo{person}{Tobias Pulls}.}
  \bibinfo{year}{2019}\natexlab{}.
\newblock \bibinfo{title}{Tor Circuit Padding Simulator}.
\newblock
\newblock
\urldef\tempurl%
\url{https://github.com/pylls/circpad-sim}
\showURL{%
\tempurl}
\newblock
\shownote{Accessed: 2022-7-14}.


\bibitem[Pulls(2022)]%
        {machines}
\bibfield{author}{\bibinfo{person}{Tobias Pulls}.}
  \bibinfo{year}{2022}\natexlab{}.
\newblock \bibinfo{title}{Towards Effective and Efficient Padding Machines for
  Tor}.  (\bibinfo{year}{2022}).
\newblock
\urldef\tempurl%
\url{https://arxiv.org/pdf/2011.13471.pdf}
\showURL{%
\tempurl}


\bibitem[Pulls and Dahlberg(2020)]%
        {oracle}
\bibfield{author}{\bibinfo{person}{Tobias Pulls} {and} \bibinfo{person}{Rasmus
  Dahlberg}.} \bibinfo{year}{2020}\natexlab{}.
\newblock \showarticletitle{Website Fingerprinting with Website Oracles}.
\newblock \bibinfo{journal}{\emph{Proceedings on Privacy Enhancing
  Technologies}}  \bibinfo{volume}{2020} (\bibinfo{date}{01}
  \bibinfo{year}{2020}), \bibinfo{pages}{235--255}.
\newblock
\urldef\tempurl%
\url{https://doi.org/10.2478/popets-2020-0013}
\showDOI{\tempurl}


\bibitem[Rahman et~al\mbox{.}(2021)]%
        {Rahman2021}
\bibfield{author}{\bibinfo{person}{Mohammad~Saidur Rahman},
  \bibinfo{person}{Mohsen Imani}, \bibinfo{person}{Nate Mathews}, {and}
  \bibinfo{person}{Matthew~K. Wright}.} \bibinfo{year}{2021}\natexlab{}.
\newblock \showarticletitle{Mockingbird: Defending Against Deep-Learning-Based
  Website Fingerprinting Attacks With Adversarial Traces}.
\newblock \bibinfo{journal}{\emph{IEEE Transactions on Information Forensics
  and Security}}  \bibinfo{volume}{16} (\bibinfo{year}{2021}),
  \bibinfo{pages}{1594--1609}.
\newblock


\bibitem[Rahman et~al\mbox{.}(2020)]%
        {Rahman2019}
\bibfield{author}{\bibinfo{person}{Mohammad~Saidur Rahman},
  \bibinfo{person}{Payap Sirinam}, \bibinfo{person}{Nate Matthews},
  \bibinfo{person}{Kantha~Girish Gangadhara}, {and} \bibinfo{person}{Matthew
  Wright}.} \bibinfo{year}{2020}\natexlab{}.
\newblock \showarticletitle{{Tik-Tok: The Utility of Packet Timing in Website
  Fingerprinting Attacks}}.
\newblock \bibinfo{journal}{\emph{Proceedings of Privacy Enhancing
  Technologies}} (\bibinfo{year}{2020}).
\newblock
\urldef\tempurl%
\url{https://petsymposium.org/popets/2020/popets-2020-0043.pdf}
\showURL{%
\tempurl}


\bibitem[Raymond(2001)]%
        {Raymond2001}
\bibfield{author}{\bibinfo{person}{Jean-Fran{\c{c}}ois Raymond}.}
  \bibinfo{year}{2001}\natexlab{}.
\newblock \bibinfo{booktitle}{\emph{Traffic Analysis: Protocols, Attacks,
  Design Issues, and Open Problems}}.
\newblock \bibinfo{publisher}{Springer Berlin Heidelberg},
  \bibinfo{address}{Berlin, Heidelberg}, \bibinfo{pages}{10--29}.
\newblock
\showISBNx{978-3-540-44702-3}
\urldef\tempurl%
\url{https://doi.org/10.1007/3-540-44702-4_2}
\showDOI{\tempurl}


\bibitem[Reed et~al\mbox{.}(1998)]%
        {onions}
\bibfield{author}{\bibinfo{person}{M.G. Reed}, \bibinfo{person}{P.F. Syverson},
  {and} \bibinfo{person}{D.M. Goldschlag}.} \bibinfo{year}{1998}\natexlab{}.
\newblock \showarticletitle{Anonymous connections and onion routing}.
\newblock \bibinfo{journal}{\emph{IEEE Journal on Selected Areas in
  Communications}} \bibinfo{volume}{16}, \bibinfo{number}{4}
  (\bibinfo{year}{1998}), \bibinfo{pages}{482--494}.
\newblock
\urldef\tempurl%
\url{https://doi.org/10.1109/49.668972}
\showDOI{\tempurl}


\bibitem[Reference(2023)]%
        {geomspace}
\bibfield{author}{\bibinfo{person}{NumPy Reference}.}
  \bibinfo{year}{2023}\natexlab{}.
\newblock \bibinfo{title}{numpy.geomspace}.
\newblock
\newblock
\urldef\tempurl%
\url{https://numpy.org/doc/stable/reference/generated/numpy.geomspace.html}
\showURL{%
\tempurl}
\newblock
\shownote{Accessed: 2023-7-07}.


\bibitem[Rimmer et~al\mbox{.}(2018)]%
        {Rimmer2018}
\bibfield{author}{\bibinfo{person}{Vera Rimmer}, \bibinfo{person}{Davy
  Preuveneers}, \bibinfo{person}{Marc Juarez}, \bibinfo{person}{Tom~Van
  Goethem}, {and} \bibinfo{person}{Wouter Joosen}.}
  \bibinfo{year}{2018}\natexlab{}.
\newblock \showarticletitle{{Automated Website Fingerprinting through Deep
  Learning}}.
\newblock  (\bibinfo{year}{2018}).
\newblock
\showISBNx{1891562495}
\urldef\tempurl%
\url{https://doi.org/10.14722/ndss.2018.23105}
\showDOI{\tempurl}


\bibitem[Shan et~al\mbox{.}(2021)]%
        {dolos}
\bibfield{author}{\bibinfo{person}{Shawn Shan}, \bibinfo{person}{Arjun~Nitin
  Bhagoji}, \bibinfo{person}{Haitao Zheng}, {and} \bibinfo{person}{Ben~Y.
  Zhao}.} \bibinfo{year}{2021}\natexlab{}.
\newblock \showarticletitle{Patch-Based Defenses against Web Fingerprinting
  Attacks}. In \bibinfo{booktitle}{\emph{Proceedings of the 14th ACM Workshop
  on Artificial Intelligence and Security}} (Virtual Event, Republic of Korea)
  \emph{(\bibinfo{series}{AISec '21})}. \bibinfo{publisher}{Association for
  Computing Machinery}, \bibinfo{address}{New York, NY, USA},
  \bibinfo{pages}{97–109}.
\newblock
\showISBNx{9781450386579}
\urldef\tempurl%
\url{https://doi.org/10.1145/3474369.3486875}
\showDOI{\tempurl}


\bibitem[Shen et~al\mbox{.}(2023)]%
        {robust}
\bibfield{author}{\bibinfo{person}{Meng Shen}, \bibinfo{person}{Kexin Ji},
  \bibinfo{person}{Zhenbo Gao}, \bibinfo{person}{Qi Li},
  \bibinfo{person}{Liehuang Zhu}, {and} \bibinfo{person}{Ke Xu}.}
  \bibinfo{year}{2023}\natexlab{}.
\newblock \showarticletitle{Subverting Website Fingerprinting Defenses with
  Robust Traffic Representation}. In \bibinfo{booktitle}{\emph{32th USENIX
  Security Symposium (USENIX Security 23)}}. \bibinfo{publisher}{USENIX
  Association}.
\newblock
\urldef\tempurl%
\url{https://www.usenix.org/conference/usenixsecurity23/presentation/shenmeng}
\showURL{%
\tempurl}


\bibitem[Shmatikov and Wang(2006)]%
        {schmatikov2006}
\bibfield{author}{\bibinfo{person}{Vitaly Shmatikov} {and}
  \bibinfo{person}{Ming-Hsiu Wang}.} \bibinfo{year}{2006}\natexlab{}.
\newblock \showarticletitle{Timing Analysis in Low-Latency Mix Networks:
  Attacks and Defenses}. In \bibinfo{booktitle}{\emph{Computer Security --
  ESORICS 2006}}, \bibfield{editor}{\bibinfo{person}{Dieter Gollmann},
  \bibinfo{person}{Jan Meier}, {and} \bibinfo{person}{Andrei Sabelfeld}}
  (Eds.). \bibinfo{publisher}{Springer Berlin Heidelberg},
  \bibinfo{address}{Berlin, Heidelberg}, \bibinfo{pages}{18--33}.
\newblock
\showISBNx{978-3-540-44605-7}


\bibitem[Sirinam et~al\mbox{.}(2018)]%
        {Sirinam2018}
\bibfield{author}{\bibinfo{person}{Payap Sirinam}, \bibinfo{person}{Marc
  Juarez}, \bibinfo{person}{Mohsen Imani}, {and} \bibinfo{person}{Matthew
  Wright}.} \bibinfo{year}{2018}\natexlab{}.
\newblock \showarticletitle{{Deep fingerprinting: Undermining website
  fingerprinting defenses with deep learning}}.
\newblock \bibinfo{journal}{\emph{Proceedings of the ACM Conference on Computer
  and Communications Security}} (\bibinfo{year}{2018}),
  \bibinfo{pages}{1928--1943}.
\newblock
\showISBNx{9781450356930}
\showISSN{15437221}
\urldef\tempurl%
\url{https://doi.org/10.1145/3243734.3243768}
\showDOI{\tempurl}


\bibitem[Sirinam et~al\mbox{.}(2019)]%
        {triplet}
\bibfield{author}{\bibinfo{person}{Payap Sirinam}, \bibinfo{person}{Nate
  Mathews}, \bibinfo{person}{Mohammad~Saidur Rahman}, {and}
  \bibinfo{person}{Matthew Wright}.} \bibinfo{year}{2019}\natexlab{}.
\newblock \showarticletitle{Triplet Fingerprinting: More Practical and Portable
  Website Fingerprinting with N-Shot Learning}. In
  \bibinfo{booktitle}{\emph{Proceedings of the 2019 ACM SIGSAC Conference on
  Computer and Communications Security}} (London, United Kingdom)
  \emph{(\bibinfo{series}{CCS '19})}. \bibinfo{publisher}{Association for
  Computing Machinery}, \bibinfo{address}{New York, NY, USA},
  \bibinfo{pages}{1131–1148}.
\newblock
\showISBNx{9781450367479}
\urldef\tempurl%
\url{https://doi.org/10.1145/3319535.3354217}
\showDOI{\tempurl}


\bibitem[SPIN-UMass(2022)]%
        {bapcode}
\bibfield{author}{\bibinfo{person}{SPIN-UMass}.}
  \bibinfo{year}{2022}\natexlab{}.
\newblock \bibinfo{title}{BLANKET}.
\newblock
\newblock
\urldef\tempurl%
\url{https://github.com/SPIN-UMass/BLANKET}
\showURL{%
\tempurl}
\newblock
\shownote{Accessed: 2022-7-14}.


\bibitem[Sun et~al\mbox{.}(2017)]%
        {sun2017}
\bibfield{author}{\bibinfo{person}{Y. Sun}, \bibinfo{person}{A. Edmundson},
  \bibinfo{person}{N. Feamster}, \bibinfo{person}{M. Chiang}, {and}
  \bibinfo{person}{P. Mittal}.} \bibinfo{year}{2017}\natexlab{}.
\newblock \showarticletitle{Counter-RAPTOR: Safeguarding Tor Against Active
  Routing Attacks}. In \bibinfo{booktitle}{\emph{2017 IEEE Symposium on
  Security and Privacy (SP)}}. \bibinfo{publisher}{IEEE Computer Society},
  \bibinfo{address}{Los Alamitos, CA, USA}, \bibinfo{pages}{977--992}.
\newblock
\showISSN{2375-1207}
\urldef\tempurl%
\url{https://doi.org/10.1109/SP.2017.34}
\showDOI{\tempurl}


\bibitem[Sun et~al\mbox{.}(2015)]%
        {raptor}
\bibfield{author}{\bibinfo{person}{Yixin Sun}, \bibinfo{person}{Anne
  Edmundson}, \bibinfo{person}{Laurent Vanbever}, \bibinfo{person}{Oscar Li},
  \bibinfo{person}{Jennifer Rexford}, \bibinfo{person}{Mung Chiang}, {and}
  \bibinfo{person}{Prateek Mittal}.} \bibinfo{year}{2015}\natexlab{}.
\newblock \showarticletitle{{RAPTOR}: Routing Attacks on Privacy in Tor}. In
  \bibinfo{booktitle}{\emph{24th USENIX Security Symposium (USENIX Security
  15)}}. \bibinfo{publisher}{USENIX Association}, \bibinfo{address}{Washington,
  D.C.}, \bibinfo{pages}{271--286}.
\newblock
\showISBNx{978-1-939133-11-3}
\urldef\tempurl%
\url{https://www.usenix.org/conference/usenixsecurity15/technical-sessions/presentation/sun}
\showURL{%
\tempurl}


\bibitem[Tan et~al\mbox{.}(2016)]%
        {Tan2016}
\bibfield{author}{\bibinfo{person}{Henry Tan}, \bibinfo{person}{Michael~E.
  Sherr}, {and} \bibinfo{person}{Wenchao Zhou}.}
  \bibinfo{year}{2016}\natexlab{}.
\newblock \showarticletitle{Data-plane Defenses against Routing Attacks on
  Tor}.
\newblock \bibinfo{journal}{\emph{Proceedings on Privacy Enhancing
  Technologies}}  \bibinfo{volume}{2016} (\bibinfo{year}{2016}),
  \bibinfo{pages}{276 -- 293}.
\newblock


\bibitem[Wang et~al\mbox{.}(2014)]%
        {Wang2014a}
\bibfield{author}{\bibinfo{person}{Tao Wang}, \bibinfo{person}{Xiang Cai},
  \bibinfo{person}{Rishab Nithyanand}, \bibinfo{person}{Rob Johnson}, {and}
  \bibinfo{person}{Ian Goldberg}.} \bibinfo{year}{2014}\natexlab{}.
\newblock \showarticletitle{{Effective Attacks and Provable Defenses for
  Website Fingerprinting}}.
\newblock \bibinfo{journal}{\emph{23rd USENIX Security Symposium (USENIX
  Security 14)}} (\bibinfo{year}{2014}), \bibinfo{pages}{143--157}.
\newblock
\showISBNx{978-1-931971-15-7}
\urldef\tempurl%
\url{https://www.usenix.org/conference/usenixsecurity14/technical-sessions/presentation/wang_tao}
\showURL{%
\tempurl}


\bibitem[Wang and Goldberg(2013)]%
        {Wang2013}
\bibfield{author}{\bibinfo{person}{Tao Wang} {and} \bibinfo{person}{Ian
  Goldberg}.} \bibinfo{year}{2013}\natexlab{}.
\newblock \showarticletitle{{Improved website fingerprinting on Tor}}.
\newblock \bibinfo{journal}{\emph{Proceedings of the ACM Conference on Computer
  and Communications Security}} (\bibinfo{year}{2013}),
  \bibinfo{pages}{201--212}.
\newblock
\showISBNx{9781450324854}
\showISSN{15437221}
\urldef\tempurl%
\url{https://doi.org/10.1145/2517840.2517851}
\showDOI{\tempurl}


\bibitem[Wang and Goldberg(2016)]%
        {Wang2016}
\bibfield{author}{\bibinfo{person}{Tao Wang} {and} \bibinfo{person}{Ian
  Goldberg}.} \bibinfo{year}{2016}\natexlab{}.
\newblock \showarticletitle{{On Realistically Attacking Tor with Website
  Fingerprinting}}.
\newblock \bibinfo{journal}{\emph{Proceedings on Privacy Enhancing
  Technologies}} \bibinfo{volume}{2016}, \bibinfo{number}{4}
  (\bibinfo{year}{2016}), \bibinfo{pages}{21--36}.
\newblock
\urldef\tempurl%
\url{https://doi.org/10.1515/popets-2016-0027}
\showDOI{\tempurl}


\bibitem[Wang and Goldberg(2017)]%
        {Wang2017}
\bibfield{author}{\bibinfo{person}{Tao Wang} {and} \bibinfo{person}{Ian
  Goldberg}.} \bibinfo{year}{2017}\natexlab{}.
\newblock \showarticletitle{Walkie-Talkie: An Efficient Defense against Passive
  Website Fingerprinting Attacks}. In \bibinfo{booktitle}{\emph{Proceedings of
  the 26th USENIX Conference on Security Symposium}} (Vancouver, BC, Canada)
  \emph{(\bibinfo{series}{SEC'17})}. \bibinfo{publisher}{USENIX Association},
  \bibinfo{address}{USA}, \bibinfo{pages}{1375–1390}.
\newblock
\showISBNx{9781931971409}


\bibitem[Wang et~al\mbox{.}(2002)]%
        {Wang2002}
\bibfield{author}{\bibinfo{person}{Xinyuan Wang}, \bibinfo{person}{Douglas~S.
  Reeves}, {and} \bibinfo{person}{S.~Felix Wu}.}
  \bibinfo{year}{2002}\natexlab{}.
\newblock \showarticletitle{Inter-Packet Delay Based Correlation for Tracing
  Encrypted Connections through Stepping Stones}. In
  \bibinfo{booktitle}{\emph{Computer Security --- ESORICS 2002}},
  \bibfield{editor}{\bibinfo{person}{Dieter Gollmann},
  \bibinfo{person}{G{\"u}nther Karjoth}, {and} \bibinfo{person}{Michael
  Waidner}} (Eds.). \bibinfo{publisher}{Springer Berlin Heidelberg},
  \bibinfo{address}{Berlin, Heidelberg}, \bibinfo{pages}{244--263}.
\newblock
\showISBNx{978-3-540-45853-1}


\bibitem[webfp(2017)]%
        {crawler}
\bibfield{author}{\bibinfo{person}{webfp}.} \bibinfo{year}{2017}\natexlab{}.
\newblock \bibinfo{title}{Tor Browser Crawler}.
\newblock
  \bibinfo{howpublished}{\url{https://github.com/webfp/tor-browser-crawler}}.
\newblock
\newblock
\shownote{Accessed: 2021-11-17}.


\bibitem[Witwer et~al\mbox{.}(2022)]%
        {paddingonly}
\bibfield{author}{\bibinfo{person}{Ethan Witwer}, \bibinfo{person}{James~K.
  Holland}, {and} \bibinfo{person}{Nicholas Hopper}.}
  \bibinfo{year}{2022}\natexlab{}.
\newblock \showarticletitle{Padding-Only Defenses Add Delay in Tor}. In
  \bibinfo{booktitle}{\emph{Proceedings of the 21st Workshop on Privacy in the
  Electronic Society}} (Los Angeles, CA, USA)
  \emph{(\bibinfo{series}{WPES'22})}. \bibinfo{publisher}{Association for
  Computing Machinery}, \bibinfo{address}{New York, NY, USA},
  \bibinfo{pages}{29–33}.
\newblock
\showISBNx{9781450398732}
\urldef\tempurl%
\url{https://doi.org/10.1145/3559613.3563207}
\showDOI{\tempurl}


\bibitem[Wright et~al\mbox{.}(2009)]%
        {Wright2009}
\bibfield{author}{\bibinfo{person}{Charles~V Wright}, \bibinfo{person}{Scott~E
  Coull}, {and} \bibinfo{person}{Fabian Monrose}.}
  \bibinfo{year}{2009}\natexlab{}.
\newblock \bibinfo{booktitle}{\emph{{Traffic Morphing: An Efficient Defense
  Against Statistical Traffic Analysis}}}.
\newblock \bibinfo{type}{{T}echnical {R}eport}.
\newblock


\bibitem[Yuan et~al\mbox{.}(2019)]%
        {dnn}
\bibfield{author}{\bibinfo{person}{Xiaoyong Yuan}, \bibinfo{person}{Pan He},
  \bibinfo{person}{Qile Zhu}, {and} \bibinfo{person}{Xiaolin Li}.}
  \bibinfo{year}{2019}\natexlab{}.
\newblock \showarticletitle{Adversarial Examples: Attacks and Defenses for Deep
  Learning}.
\newblock \bibinfo{journal}{\emph{IEEE Transactions on Neural Networks and
  Learning Systems}} \bibinfo{volume}{30}, \bibinfo{number}{9}
  (\bibinfo{year}{2019}), \bibinfo{pages}{2805--2824}.
\newblock
\urldef\tempurl%
\url{https://doi.org/10.1109/TNNLS.2018.2886017}
\showDOI{\tempurl}


\bibitem[Zhu et~al\mbox{.}(2005)]%
        {ye2005}
\bibfield{author}{\bibinfo{person}{Ye Zhu}, \bibinfo{person}{Xinwen Fu},
  \bibinfo{person}{Bryan Graham}, \bibinfo{person}{Riccardo Bettati}, {and}
  \bibinfo{person}{Wei Zhao}.} \bibinfo{year}{2005}\natexlab{}.
\newblock \showarticletitle{On Flow Correlation Attacks and Countermeasures in
  Mix Networks}. In \bibinfo{booktitle}{\emph{Privacy Enhancing Technologies}},
  \bibfield{editor}{\bibinfo{person}{David Martin} {and}
  \bibinfo{person}{Andrei Serjantov}} (Eds.). \bibinfo{publisher}{Springer
  Berlin Heidelberg}, \bibinfo{address}{Berlin, Heidelberg},
  \bibinfo{pages}{207--225}.
\newblock
\showISBNx{978-3-540-31960-3}


\end{thebibliography}

\appendix

\section{Implementation Details}

\subsection{Model Architecture}
\label{ssec:architecture}
\begin{figure}[!ht]
  \caption{}
  \centering
  \includegraphics[scale=.9]{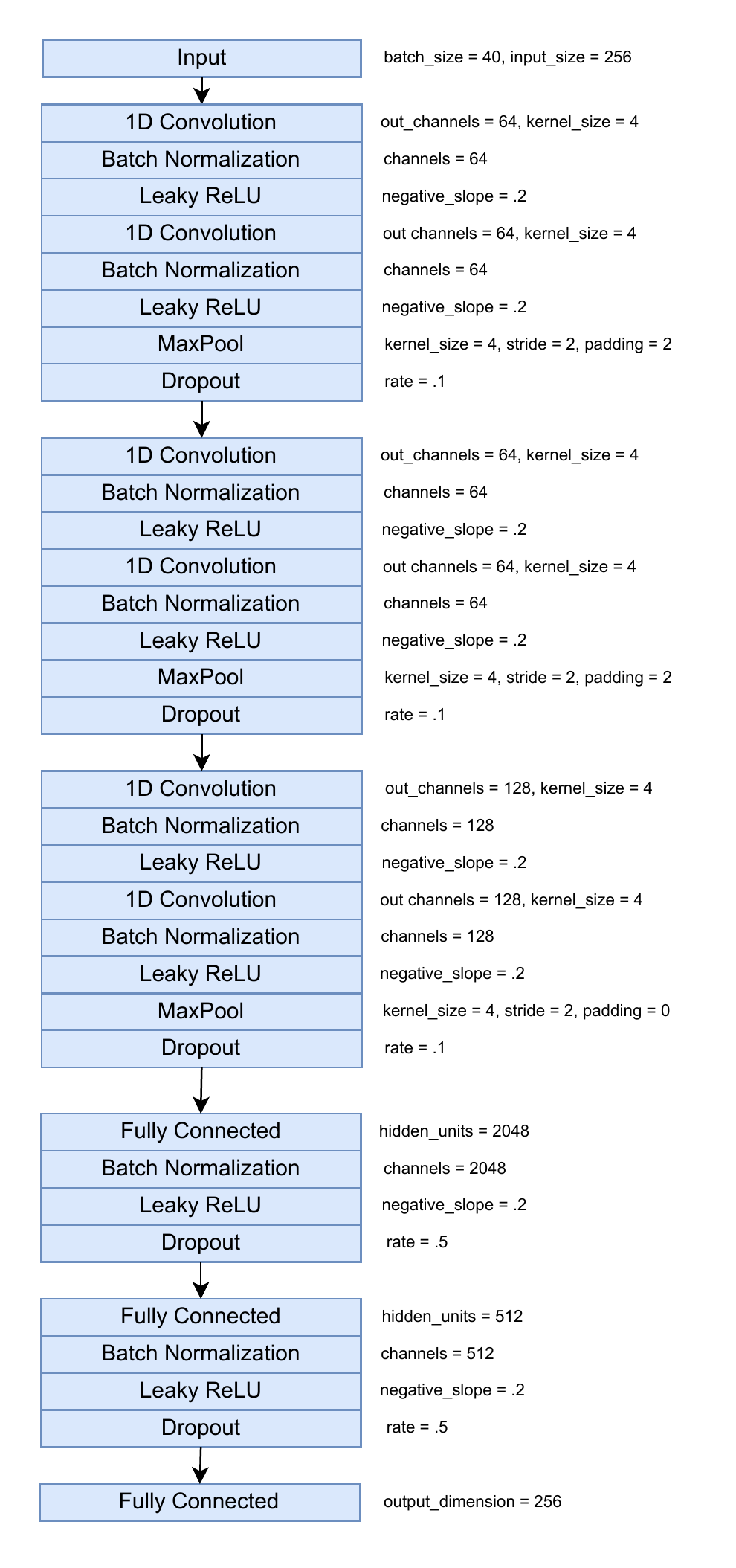}
  \label{fig:cnn}
\end{figure}

The discriminator and embedder in the WF setting use the architecture shown in Figure \ref{fig:cnn}. This CNN architecture is based on the Deep Fingerprinting architecture with alterations so that it uses a tensor of size 256 as input and output. We also find that model performance is improved with Leaky ReLU rather than ReLU activation. 

\begin{figure*}
  \caption{}
  \centering
  \includegraphics[scale=.9]{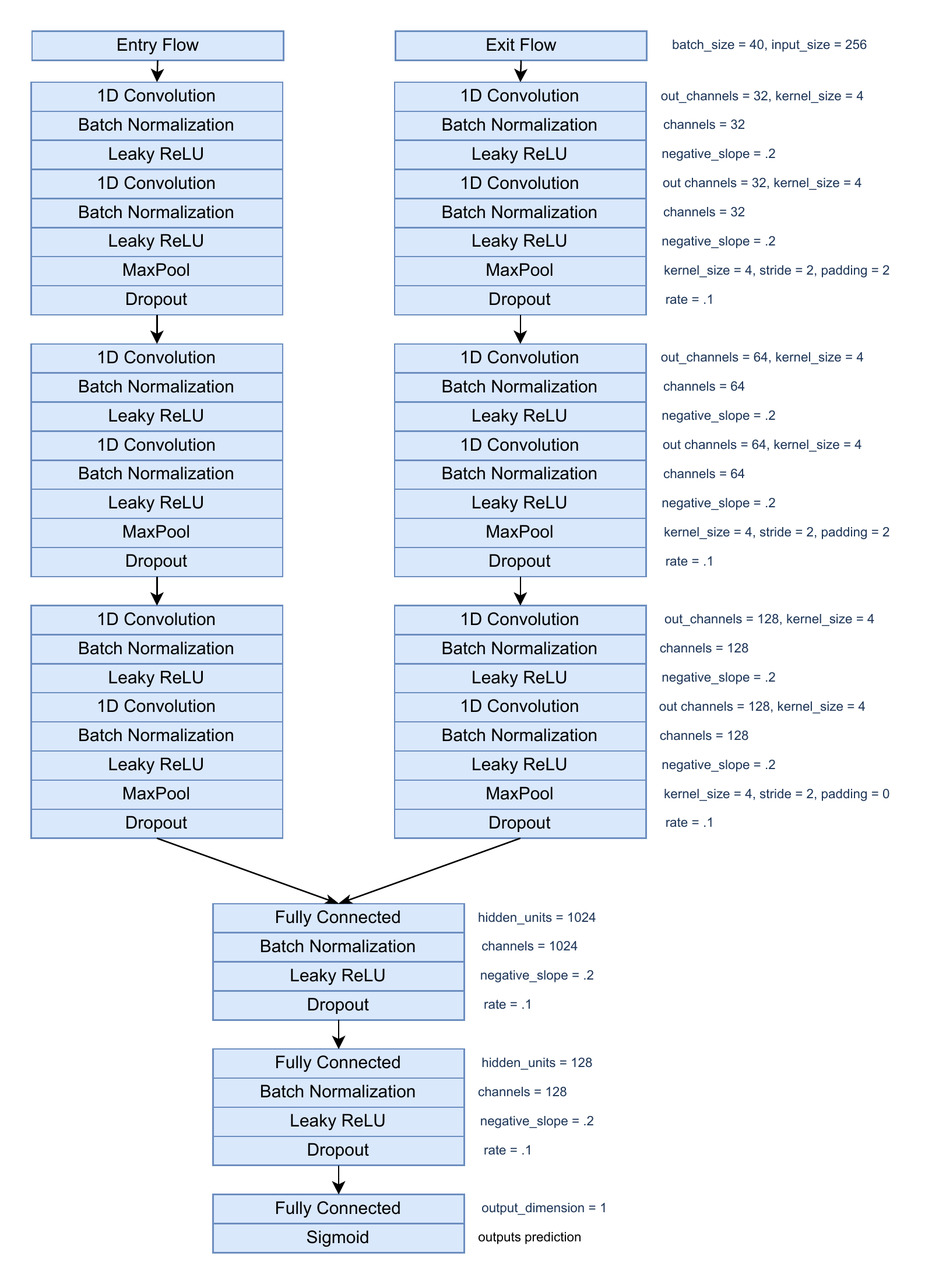}
  \label{fig:fc_cnn}
\end{figure*}

The discriminator in the FC setting is shown in Figure \ref{fig:fc_cnn}. Note that the outputs after initially processing the two flows are concatenated and then passed to the final layers. The output of the network is a prediction about whether the two flows are correlated where an output greater than .5 predicts correlation. 

Note that the input of the generator, shown in Figure \ref{fig:lstm_unrolled}, is the concatenation of random noise and the volume of real traffic binned \textit{at that time step}, and that the hidden state at each time step is used as the input to the fully connected layer. The generator LSTM has a hidden size of 128, an input size of 32, and is made up of a single layer. 

\subsection{Training and Model Distribution}

In the WF setting, we train the embedder, discriminator, and generator using the Adam optimizer with a learning rate of $.0001$. The embedder uses PyTorch's learning rate scheduler ReduceLROnPlateau with minimum a learning rate of $.000001$. The dimensionality of the trace embeddings is 256 and the batch size is 40. The embedder's triplet loss uses Euclidean distance with a margin of 1. We train the discriminator-generator pair for 90 epochs. 

The FC setting uses the same batch size and optimizer settings but trains the discriminator-generator pair for 60 epochs. We trained the models on an Nvidia GeForce RTX 3090 and find that it takes about an hour to train WF-DeTorrent and FC-DeTorrent. However, simulating DeTorrent on the chosen dataset and evaluating defense performance generally takes 3 to 4 more hours, making hyperparameter optimization prohibitively time-consuming. While using a dataset of 19,000 traces, the training process uses a maximum of 3 GiB of GPU memory. 

With the trained models, we are able to use the GPU to generate a defended trace in about 140ms, making it possible to generate a defended dataset relatively quickly. The saved defense generator model is quite small at only 358 KiB as it has only about 86,400 trainable parameters. 

One implementation challenge for DeTorrent is distributing the models. Because the upload padding is quite simple, only the generator needs to be distributed to bridges or relays. Our proposal for model distribution is similar to the one presented for the Surakav defense \cite{surakav}, which is that the semi-trusted Tor directory servers train, distribute, and occasionally update the defense generators. Given the small size of the DeTorrent generator, we don't expect the additional bandwidth to excessively burden the directory servers. 

\section{Estimating Traffic Rate}
\label{ssec:timing_details}
To determine the number of dummy upload packets to send so that the total upload rate is one-fifth of the download packet rate, we need to maintain accurate and frequently updated rate estimations for the rates of real upload and download traffic. However, this is nontrivial, as we'll need to find the rate of occurrence of an irregular event that is characterized by infrequent but large `bursts' of activity. We also need responsive estimates in the sense that they quickly update to match the sending rate of traffic (rather than being lagging indicators) and we need to be able to compute an accurate average \textit{between} packets. In other words, if an upload packet hasn't been sent in a while, then we should take that into account while computing an average. Lastly, for practical performance reasons, we would like to avoid searching for or storing information about previous traffic timing, as we'll have to compute the packet sending rate very frequently.

The approach described by Ilmari Karonen \cite{packetrate} first considers that we can model the true frequency over the whole time period as the integral of Dirac delta functions centered at the occurrence of the event divided by an integral of a constant (1) over time where $\lambda_{packet\_rate}$ represents the estimated packet sending rate:

\[\lambda_{packet\_rate} = \frac{\int \delta_{sent} (\tau) d\tau}{\int 1 d\tau}\]

However, we want an estimate of the \textit{current} packet sending rate rather than the rate over the time period. So, we add a weighting function to emphasize the more recent packet sending:

\[\lambda_{recent\_rate} = \frac{\int \delta_{sent} (\tau) w(\tau) d\tau}{\int w(\tau) d\tau}\]

One reasonable choice of weighting function is to make it exponential, such as $w(\tau) = e^{k(\tau - t)}$ for a decay rate $k$. This also allows us to make the following simplification: 

\[ \lambda_{recent\_rate} = \frac{k \sum_{i=0}^{i=N} e^{k(t_i - t)}}{1-e^{k(t_0 - t)}} \approx k \sum_{i=0}^{i=N} e^{k(t_i - t)}\]\\ for large $t - t_0$\\

To avoid having to save all of the event times, we can simply store the outcome of the most recent calculation and update it as follows:

\[ \lambda_{recent\_rate}(t) = e^{k(t' - t)} \lambda_{recent}(t')\] where $t$ is the current time and $t'$ is the time of the previous rate calculation.

When a new packet arrives, we'll account for it by updating as follows: 

\[ \lambda_{recent\_rate}(t) = k + e^{k(t' - t)} \lambda_{recent}(t')\] 

Thus, we have a method of calculating the sending rate that can be tuned to be responsive to recent bursts, requires little information storage or computation, can handle the irregular nature of internet traffic, and allows us to accurately compute the rate both alongside and between the networking events. While the choice of $k$ may vary, we find that $k=1$ best fits the datasets used in this paper. 

\end{document}